\documentclass[twocolumn]{aastex6}

\usepackage{amsmath, natbib}

\newcommand{\ssfr}{\Sigma_{\rm SFR}}
\newcommand{\msun}{M_\odot}
\newcommand{\mstar}{m_\star}

\newcommand{\lrot}{\lambda_{\rm rot}}
\newcommand{\lj}{\lambda_{\rm Jeans}}

\newcommand{\tdyn}{t_{\rm dyn}}
\newcommand{\torb}{t_{\rm orb}}
\newcommand{\tcool}{t_{\rm cool}}

\newcommand{\tff}{t_{\rm ff}}

\newcommand{\rhogas}{\rho_{\rm gas}}
\newcommand{\sstar}{\Sigma_\star}

\newcommand{\sgas}{\Sigma_{\rm gas}}

\newcommand{\mgas}{M_{\rm gas}}

\newcommand{\cs}{c_{\rm s}}

\newcommand{\vturb}{v_{\rm turb}}
\newcommand{\rcyl}{R_{90}}
\newcommand{\zcyl}{z_{90}}

\begin{document}


\title{Unveiling the role of galactic rotation on star formation}


\author{Jos\'e Utreras\altaffilmark{1}, Fernando Becerra\altaffilmark{2}, Andr{\'e}s Escala\altaffilmark{1}}
\affil{$^1$Departamento de Astronom\'ia, Universidad de Chile, Casilla 36-D, Santiago, Chile\\
$^2$Harvard-Smithsonian Center for Astrophysics, 60 Garden Street, Cambridge, MA 02138, USA}
\shorttitle{Unveiling the role of galactic rotation on star formation}
\shortauthors{Utreras et al.}

\begin{abstract}

We study the star formation process at galactic scales and the role of rotation through numerical simulations of spiral and starburst galaxies using the adaptive mesh refinement code Enzo. We focus on the study of three integrated star formation laws found in the literature: the Kennicutt-Schmidt (KS) and Silk-Elmegreen (SE) laws, and the dimensionally homogeneous equation proposed by \citet{Escala_15} $\ssfr \propto \sqrt{G/L}\sgas^{1.5}$. We show that using the last we take into account the effects of the integration along the line of sight and find a unique regime of star formation for both types of galaxies, suppressing the observed bi-modality of the KS law. We find that the efficiencies displayed by our simulations are anti-correlated with the angular velocity of the disk $\Omega$ for the three laws studied in this work. Finally, we show that the dimensionless efficiency of star formation is well represented by an exponentially decreasing function of $-1.9\Omega t_{\rm ff}^{\rm ini}$, where $t_{\rm ff}^{\rm  ini}$ is the initial free-fall time. This leads to a unique galactic star formation relation which reduces the scatter of the bi-modal KS, SE, and \citet{Escala_15} relations by 43\%, 43\%, and 35\% respectively.

\end{abstract}


\section{INTRODUCTION}
\label{sec:intro}
Knowing how efficiently galaxies form their stars is fundamental to understanding the evolution of our universe. Unfortunately, star formation involves a wide range of spatial and temporal scales combined with a large number of physical processes which are usually dynamically coupled and in the nonlinear regime, thus a comprehensive understanding of how stars form has remained elusive to astronomers. The typical picture for star formation consists of dense molecular clouds collapsing due to their own gravity, while resisted by thermal pressure, turbulence, rotational shear, and magnetic fields, among others \citep{Martin_01,Seigar_05,McKee_07,Padoan_14}. Studies have shown that stars are most likely created from molecular clouds on scales of a few parsecs, and that feedback produced by their subsequent explosion as supernovae can impact their surroundings up to scales of a few kiloparsecs \citep{Ceverino_09}. The influence that stellar feedback has on galactic scales naturally leads to the assumption that star formation might be somewhat related to galactic properties. \\

One of the first attempts to find such link was performed by \citet{Schmidt_59}, who suggested a power-law relation between the volume densities of star formation rate (SFR) and gas content, based on observations of the solar neighborhood. Further observations of disk and starburst galaxies performed by \citet{Kennicutt_98} strengthened this hypothesis and found a relation of the form $\ssfr \propto \sgas^N$ with $N \approx 1.4$, the so-called Kennicutt-Schmidt (KS) relation. From this simple relation it follows that the depletion times ($t_{\rm SF}$) might be described by $t_{\rm SF}  \equiv \sgas/\ssfr \propto \sgas ^{-0.4}$ which is often associated with the free-fall time of the gas $t_{\rm ff} \propto (G \rho_{\rm gas})^{-0.5} $\citep{Madore_77,Leroy_08}.\\

However, the functional form of this relation appears not to be unique: normal spiral galaxies seem to create stars from their gas reservoirs over longer time-scales than high-redshift starburst disks at the same surface density, suggesting the existence of two different regimes in the KS relation \citep{Daddi_10, Genzel_10}. Furthermore, other works have shown variations of this law depending on the tracers used, casting doubts on how fundamental these relations are. For instance, in H$_2$-dominated centers of spiral galaxies star formation has been found to be linearly related to the density of molecular hydrogen instead of atomic hydrogen \citep{Bigiel_08}. On the other hand, observations of HI-dominated regions show that star formation is related to atomic hydrogen through the KS dependence with depletion times 1-2 orders of magnitude lower than in the inner molecular dominated disk \citep{Roy_15}. These results are compatible since the SFR grows linearly with the mass of molecular clouds \citep{Lada_10}, which are formed from diffuse gas and therefore the laws based on the total hydrogen content represent the formation of such clouds.\\

Although the KS relation is probably the most extensively studied star formation law, it is not the only effort that has been performed in this area. Several authors have proposed other formulations where the $t_{\rm SF}$ is related to a galactic time scale, such as the effective free-fall time at the mid-plane of the disk \citep{Krumholz_12}, or the orbital time $\torb =2\pi /\Omega$ known as the Silk-Elmegreen (SE) relation \citep{Silk_97,Elmegreen_97}, where $\Omega$ is the angular velocity. Unfortunately, in real galaxies both time-scales are of the same order of magnitude (at galactic scales) and are coupled through the mass content of the galaxy, which might lead to spurious relationships. In particular, a few studies show conflicting results about the role of galactic rotation: some works have claimed formation of stars through cloud collisions as the origin of the SE law \citep{Tasker_09,Suwannajak_14} while other groups have shown that collisions produce little difference \citep{Dobbs_15} in the SFR and found anti-correlations between SFR or stellar content and rotation \citep{Berta_08,Weidner_10,Davis_14,Obreschkow_15}.

Recently, \citet{Escala_15} proposed a star formation law of the form $\ssfr \propto \sqrt{\frac{G}{L}}\sgas^{3/2}$ (hereafter E15), where $L$ is a characteristic length that is related to the integration axis in observations. Among different choices for $L$, the author points out the radius $R$ of the region, the vertical scale length $h$, the Jeans length $\lj = \sqrt{\pi c_s ^2 /G\rho}$, and the largest scale not stabilized by rotation $\lrot =4\pi^2G \Sigma /\kappa^2$ \citep{Escala_08} (where $\kappa$ is the epicyclic frequency and in general $\mathbf{\Omega \lesssim \kappa \lesssim 2\Omega}$) as the main possibilities. Each one of these values allowed the author to recover previously proposed scaling relations such as $\ssfr$ with $\sgas^{3/2}$, $\sgas/\torb$, $\sgas/\tff$, and $\sgas^2/\vturb$.\\

In this paper we study the role of galactic rotation on star formation by performing three-dimensional hydrodynamic simulations of spiral and starburst galaxies which allow us to span a wide range of $\sgas$. The initial configuration of each galaxy is set by several length scales and characteristic masses of gas, stars and/or dark matter (DM), which translates into different density distributions, vertical accelerations, and rotational velocity curves. For a rotating disk, increasing the initial gas content increases the angular velocity and the vertical acceleration at the mid-plane, producing a spurious correlation between them. In addition, as the gas cools it quickly settles on the mid-plane, increasing the local gas density. This vertical compression causes the vertical acceleration to be highly correlated with the gas distribution after the first vertical collapse. On the other hand, the angular velocity is less sensitive to the vertical distribution of gas and remains nearly constant. Therefore, it is easier to study the role of $\Omega$ since it is less affected by the process of collapse and star formation. To remove the correlation between rotation and vertical acceleration we fix the initial gas content and make use of external potentials to obtain different angular velocities. The mass of these potentials is distributed on larger vertical scale-lengths, like spherically symmetric potentials, and thus does not change the vertical distribution of gas significantly which is going to be given mostly by gas self-gravity. This procedure allows us to nearly isolate the effects of rotation.\\

The paper is organized as follows. In Section 2 we introduce the simulation setup, describe the code and give details about the initialization for both spiral and starburst galaxies. In Section 3 we display the resulting structures of the simulations. We then analyze the simulations and present the results for the star formation laws in Section 4. Section 5 presents a discussion of the results, and we conclude in Section 6.

\section{THE MODELS}
\label{sec:models}

We use the hydrodynamic adaptive mesh refinement (AMR) code Enzo\footnote{http://enzo-project.org} \citep{Bryan_14}. Its Eulerian nature based on the structured AMR algorithm of \citet{Berger_89} allows different levels of refinement in different regions of space, which focuses the computation effort in regions where it is most useful. This is achieved by subdividing the ``parent" or root grids into smaller or ``child" grids, and repeating the same process when a child grid becomes itself a parent grid. Using this approach, the initial uniform grid covering the simulation box ends up turning into a nested structure of grids, where the smaller the grid size, the higher the resolution.\\

A three-dimensional version of the ZEUS hydrodynamical code \citep{Stone_92} is used to evolve the hydrodynamic equations of the gas. To ensure that the interstellar medium (ISM) will develop a multi-phase structure, a cooling function is used to calculate the energy lost by radiation down to 300 K. We use the cooling curves of \citet{Sarazin_87} for $T>10^4$ K, and \citet{Rosen_95} for $300$ K$<T<10^4$ K.\\

For this work we have performed two different sets of simulations: spiral and starburst galaxies, which allow us to probe different time, length, and density scales, at the same time that they achieve different resolutions. The simulations are evolved in comoving coordinates in a $\Lambda$CDM universe, where we have adopted the values $\Omega _m = 0.3$, $\Omega _{\Lambda} = 0.7$, and $\rm H_0 = 67 km s^{-1} Mpc^{-1}$, whose effects are only noticeable in the spiral galaxies. We use two criteria in order to refine a given gas cell and both of them have to be fulfilled: refinement by baryon mass and Jeans length. Baryon mass refinement is applied if the mass of the cell is $\delta$ times greater than $\bar{\rho}(\Delta x_{\rm root})^3 2^{0.5 l}$, where $\bar{\rho}$ is the average density, $\Delta x _{\rm root}$ is the cell size of the root grid and $l$ is the level of refinement. The second criterion is used to ensure that the Jeans length is at least resolved by four cells to prevent artificial fragmentation \citep{Truelove_97}. We choose $\delta = 4$ for spiral galaxies and $\delta = 100$ for starburst. For spiral galaxy simulations we choose a time interval between snapshots of $\Delta t \simeq$ 40 Myr, and $\Delta t \simeq 0.4$ Myr for starburst galaxies. The initial circular velocity for each simulation is calculated as
\begin{equation}
\label{eq:circ}
{v_c ^2 =R \frac{d\Phi_{\rm ext}}{dR} + \frac{GM_{\rm gas}}{R}}
\end{equation}
where $\Phi_{\rm ext}$ is the external potential and ${M_{\rm gas}}$ is the gaseous mass enclosed in a sphere of radius $R$. The second term in equation \ref{eq:circ} is an approximation for the rotational support needed for gas self-gravity. However, the radial acceleration is mostly given by the external potentials, making this approximation reliable.

\subsection{Spiral galaxies}
\label{subsec:galaxies}
The spiral galaxies are simulated in a box of $666h^{-1}$ kpc with periodic boundary conditions from $z=0.2$ to $z=0$. The size of the parent grid is $128^3$ and we proceed down to an additional seven sub-grids of refinement, reaching a resolution of $\rm  \sim 40 pc$, which is a reasonable resolution to resolve the interaction between star formation, stellar feedback, and the ISM \citep{Ceverino_09}. These simulations are modeled as a four-component system which includes gas, star particles, and time-independent stellar and DM potentials. In the case of gas we model it using grids, while the stellar and DM distributions are represented by external potentials which are fixed in time. Star particles form from the gas cells and are not added at the beginning of the simulations. The initial conditions for these galaxies were taken from \citet{Becerra_14}.
\newpage
\subsubsection{Gas}

The gas is a rotationally supported disk, initially described by an exponential profile in the radial direction $R$ combined with a sech$^2$ profile in the vertical direction $z$, where $R$ and $z$ are cylindrical coordinates

\begin{equation}\label{eq:1}
\rhogas(R,z)=\rho_0 {\rm exp}\left(-R/R_0\right) {\rm sech}^2\left(\frac{z}{2z_0}\right),
\end{equation}
where $R_0$ is the disk scale-length, $z_0$ is the disk scale-height and $\rho_0$ is the central density. Integrating equation \ref{eq:1} we obtain 

\begin{equation}
\rho _0=\frac{M_{\rm gas}}{8\pi z_0 R_0 ^2}
\end{equation}

Therefore the distribution of gas is fully determined by $R_0$, $z_0$ and the total gas mass $ M_{\rm gas}$. For the spiral galaxy set of simulations we use $R_0 = 3.5$ kpc, $z_0 = 0.4$ kpc and $ M_{\rm gas}= 10^{10} M_{\odot}$.

\subsubsection{Stars}
To model the stellar component of the external potential we use a Miyamoto-Nagai profile \citep{Miyamoto_Nagai_75}, which models the stellar disk and bulge of the galaxy. The potential is given by: 
\begin{equation}
\Phi _{\rm star}(R,z)= - \frac{GM_{\rm star}}{\sqrt{R^2+(a+\sqrt{z^2+b^2})^2}}
\end{equation}

We adopt the fixed values $a= 3.5$ kpc and $b=$0.2kpc. For the total stellar mass $M_{\rm star}$ we take two different values, $10^{10}  M_{\odot}$ and $10^{11}  M_{\odot}$ which would give in different contributions to the total rotation.

\subsubsection{Dark Matter}
We consider DM as an external gravitational field given by a Navarro-Frenk-White profile \citep{NFW97} which changes slightly through the evolution of the galaxies due to its dependence with the Hubble parameter $H$. This approach allows us to focus only on the gas dynamics. The NFW density profile is given by

\begin{equation}
\rho _{\rm DM}(r)=\frac{\rho _{\rm crit} \delta _c}{(r/r_s)(1+r/r_s)^2}
\end{equation}

where $r_s=r_{200}/c$ is the characteristic radius, $\rho _{\rm crit}=3H^2/8\pi G$ is the critical density, $c$ is the concentration parameter, and $\delta _c$ is given by:
\begin{equation}
\delta _c=\frac{200}{3}\frac{c^3}{[\ln(1+c)-c(1+c)]}
\end{equation}
The characteristic radius $r_{200}$ corresponds to the volume at which the mean density is 200 times the critical density
 \begin{equation}
 M_{200}=200 \rho _{\rm crit} \frac{4 \pi}{3} r_{200}^3
 \end{equation}
 
We adopt a value $c = 12$ for the concentration parameter while $M_{200}$ will take the values $10^{10}  M_{\odot}$ and $10^{11}  M{\odot}$ for the simulations with $M_{\rm star}=10^{10}  M_{\odot}$ and $10^{11} M_{\odot}$ respectively.
The simulation parameters for galaxies are summarized in Table \ref{table:all_parameters}.
\floattable
\begin{deluxetable}{ccccccc}
\tablecaption{Simulation parameters\label{table:all_parameters}}
\tablewidth{1pt}
\tablehead{
\colhead{} & \colhead{} & \colhead{} & \colhead{Spiral Galaxies} &
\colhead{} & \colhead{} & \colhead{} }
\startdata 
Run & $ M_{\rm gas} $ & $ M_{\rm star}$ & $ M_{\rm DM}$ & $t_{\rm orb}$  & $\Omega$ & $\Delta x$\\ 
Name & $({M}_{\odot})$ & $({M}_{\odot})$ & $({M}_{\odot})$ & $(\rm{Myr})$ & $(\rm{Myr} ^{-1})$&$(\rm pc)$\\ 
\hline                                
    GD1 & $1 \times 10^{10}$ & $1 \times 10^{10}$ & $1 \times 10^{10}$ & $639.9$ & $0.982\times 10^{-2}$ & $37.957$\\      
    GD2 & $1 \times 10^{10}$  & $1 \times 10^{11}$ & $1 \times 10^{11}$ & $275.5$ & $2.281 \times 10^{-2}$ & $37.957$\\
\hline   
&&&Starburst Galaxies &&&\\
\hline
Run & ${ M_{\rm gas}}$ & $ M_{\rm star}$ & $\sigma$  & $t_{\rm orb}$ & $\Omega$& $\Delta x$ \\ 
Name & $({M}_{\odot})$ & $({M}_{\odot})$ & $(\rm{ km/s})$ & $(\rm{Myr})$ & $(\rm{Myr} ^{-1})$&$(\rm pc)$\\ 
\hline                                
    SD1 & $4 \times 10^8$ & $1.24 \times 10^9$ & $100$ & $11.5$ & $0.546$ &$1.949$\\      
    SD2 & $4 \times 10^8$ & $2.10 \times 10^9$ & $130$ & $9.3$ & $0.675$ &$1.949$\\
    SD3 & $4 \times 10^8$ & $4.49 \times 10^9$ & $190$ & $6.6$ & $0.952$ &$1.949$\\
    SD4 & $4 \times 10^8$ & $6.02 \times 10^9$ & $220$ & $5.8$ & $1.083$ &$1.949$\\
\enddata
\tablecomments{ Initial masses, angular velocity, orbital time, and maximum resolution for each simulation. For spiral galaxies the table shows $M_{\rm gas}$, $M_{\rm star}$, $M_{\rm DM}$ and the orbital time at 10 kpc from the center. For starburst galaxies the table shows $\rm M_{\rm gas}$, $\rm M_{\rm star}$ and its corresponding dispersion velocity for the isothermal sphere and the orbital time at 300 pc from the center.}
\end{deluxetable}

\subsection{Starburst galaxies}
\label{subsec:nuclear_disks}

For starburst galaxies we only simulate the central nuclear disks which are initialized within a box of physical size 4 kpc with isolated boundary conditions. The size of the parent grid is $32^3$, and we proceed down to additional six levels of refinement, reaching a resolution of $\sim 2$ pc. Our initial model consists of a massive gaseous disk embedded in a stellar spheroid, which is modeled by a time-independent external potential. 

\subsubsection{Gas}
We initialize the gas as a rotationally supported disk with an $R^{-1}$ power law for the cylindrical radius and a sech$^2$ function for the vertical component:
\begin{equation}
\rhogas(R,z)=\rho_0\left(\frac{R_0}{R}\right) {\rm sech}^2\left(\frac{z}{2z_0(R)}\right),
\end{equation}
where $R_0$ is the radial scale-length, $z_0(R)$ is the height scale-length as a function of radius, and $\rho _0 $ is the central volumetric density. We choose $z_0$ to be a function of radius so the initial configuration is close to vertical equilibrium:
\begin{equation}
z_0(R)=\sqrt{\frac{\cs^2}{8\pi G \rho(R,0)}}=\frac{2}{3}\frac{\cs^2}{G\mgas}\sqrt{R R_0 ^3},
\end{equation}

where $M_{\rm gas}$ is the total gas mass and $c_{\rm s}$ is the sound speed which is initially constant. The values adopted for this set are $R_0 = 300\,{\rm pc}$ for the disk scale-length and $\mgas = 4\times 10^8 {M}_{\odot}$ for the total gas mass. Finally, the initial gaseous disk is truncated at $R_0$, and we add random density and temperature fluctuations of less than 10\%.

\subsubsection{Stars}

The stellar spheroid is modeled by means of the merge of two analytical density functions such that the resulting function shares some properties with an isothermal sphere. For $r<r_0$ we use a fourth-order polynomial, and the singular sphere solution for $r>r_0$, being $r_0 = \sqrt{9\sigma ^2/(4\pi G\rho _0)}$ the King radius \citep{Binney_Tremaine_08}, $\sigma$ the velocity dispersion and $\rho _0$ the stellar density at the center of the disk. In the case of the polynomial, we impose four constraints to determine the coefficients: the function and its first derivative (i,ii) have the same value of an isothermal sphere at the center, and (iii,iv) are continuous at $r = r_0$. Solving for the polynomial coefficients, we obtain:

\begin{equation}
\rho_{\rm star}(r) =  \rho_0\left\{
\begin{array}{cc}
\left[ 1-\frac{8}{9}\left(\frac{r}{r_0}\right)^2 -\frac{8}{9}\left(\frac{r}{r_0}\right)^3 +\left(\frac{r}{r_0}\right)^4\right]  & \text{if }r<r_0\\
\frac{2}{9}\left(\frac{r_0}{r}\right)^2  & \text{if } r\geq r_0
\end{array}\right.
\end{equation}

We perform a set of four different runs with a fixed $r_0 = 100$ pc. We vary the mass of the stellar spheroid in such a way that the velocity dispersion is $100$, $130$, $190$ and $220$ km/s. We let each simulation evolve adiabatically for $1.5 \torb$, after which radiate cooling is turned on. We do this to allow the initial perturbations to propagate through the disk and to remove any artefact produced by the initial conditions. Table \ref{table:all_parameters} summarizes the parameters used in these runs and indicates the value of the stellar mass within $300$ pc ($M_{\star}$). Here, $\torb$ corresponds to the period for a circular orbit at $r = 300$ pc. 

\subsection{Star formation and feedback}

We follow a slightly modified version of the \citet{Cen_Ostriker_92} algorithm to create a new star particle. For a cell to enter the algorithm of star formation its density has to be greater than a density threshold
\begin{equation}
n _{\rm cell} > n _{\rm thres}
\end{equation}
where $n$ is the particle number density. Due to the different ranges of densities and resolutions reached by each set of simulations, we use two different values: for spirals $n _{\rm thres}= {\rm 7.5\times 10^3  cm^{-3}}$ and $n _{\rm thres}={\rm 7.5\times 10^5 cm^{-3}}$ for starbursts. These values were chosen in order to match the observed KS relations for spiral and starburst galaxies \citep{Daddi_10}. If the density criterion is fulfilled, two additional criteria have to be satisfied to create a new star particle: (i) the velocity flow is converging and (ii) the time it takes the gas to cool is less than the time it takes to collapse:
\label{subsec:sf}

	\begin{subequations}
		\begin{equation}
		\vec{\nabla} \cdot \vec{v} <0
		\label{eq:divV}
		\end{equation}
		\begin{equation}
		\tcool < \tdyn \equiv \sqrt{\frac{3\pi}{32G\rho_{tot}}}
		\label{eq:tcooldyn}
		\end{equation}
	\end{subequations}
	
Finally a new star is created and its mass is calculated as a function of the star formation efficiency ($\varepsilon$), the gas density ($\rhogas$), and the cell volume: $\mstar=\varepsilon \rhogas \Delta x^3$, where $\varepsilon =0.01$. The main difference with the original Cen \& Ostriker algorithm is that we do not check if the cell is Jeans unstable because any cell that satisfies the density criterion is Jeans unstable if the cell temperature is below $1\times 10^5$ K. Unlike the Cen \& Ostriker algorithm, here we do not consider the efficiency per dynamical time, which means that there is no delay in star formation. After the star particles are formed 40\% of the gas content is returned slowly to the cell within $12 t_{\rm dyn}$.\\

We also include stellar feedback due to supernovae explosions, injecting energy to the surrounding gas. Although star particles are instantly formed in our simulation, these particles actually represent a group of several stars, hence the feedback energy is computed as if this group, within a single star particle, were created over a longer period of time. Quantitatively, the mass of stars formed in a cell at a time $t$ with time step $\Delta t$ is
	
$$\Delta m_{\rm stars}(t)=\mstar \int ^{\tau _1} _{\tau _0} \tau e^{-\tau}d\tau$$
 \begin{equation} 
 \Delta m_{\rm stars}(t)= 0.6\mstar[(1+\tau _0)e^{-\tau _0}-(1+\tau _1)e^{-\tau _1}]
 \end{equation}
where $\tau _0=(t-t_{\rm ct})/t_{\rm dyn}$, $\tau _1=(t+\Delta t-t_{\rm ct})/t_{\rm dyn}$ and $t_{\rm ct}$ is the creation time of the star particle. Finally a fraction of the rest mass energy $f_{SN} \Delta m_{\rm stars}c^2$, corresponding to the massive stellar population within the ``star particle", is injected to the grid cell that contains the particle as thermal energy. The fraction is chosen such that $10^{51}$ erg are injected for every $55 M_{\odot}$. 
	\begin{figure*}[!ht]
	\begin{center}
	\includegraphics[width=0.8\textwidth ,natwidth=1100,natheight=1100]{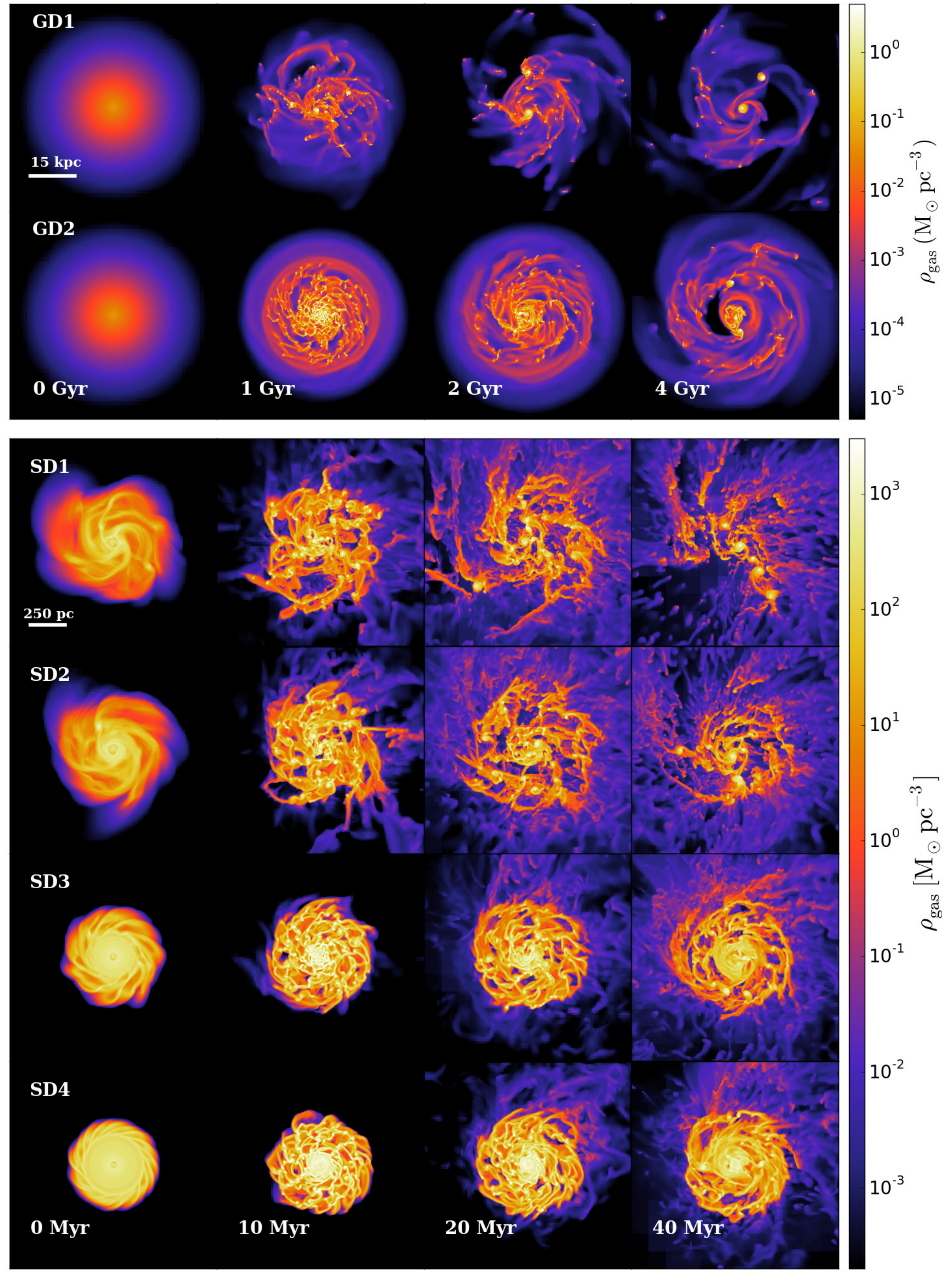}
	\caption{The panels show density-weighted projections of gas density for each simulation.  Top: runs GD1 (top) and GD2 (bottom) at times 0, 1, 2 and 4 Gyr from left to right. GD1 shows a faster departure from axial symmetry due to its weaker potential and a faster consumption of gas compared with GD2.  Bottom: Runs SD1, SD2, SD3 and SD4 from top to bottom at times 0, 10, 20 and 40 Myr from left to right, where $t=0$ Myr represents the time at which cooling and star formation is turned on. The simulation with the lowest angular velocity SD1, shows a fast departure from axial symmetry and a fast consumption of gas relative to the other starburst galaxies.}
		\label{fig:gas_galaxy}

	\end{center}
	\end{figure*}

\section{GLOBAL EVOLUTION}
\label{sec:evolution}

\subsection{Disk Structure}
We start by looking at the structure evolution of each galaxy which is going to be affected mainly by rotation, turbulence, and self-gravity. Figure \ref{fig:gas_galaxy} shows the time evolution of spiral and starburst galaxies respectively. Snapshots represent density-weighted projections of gas density at 0, 1, 2 and 4 Gyr for spirals and at 0, 10, 20 and 40 Myr for starbursts. These projections show filamentary structures and high-density blobs indicating small-scale instabilities growing in each disk. The latter is more evident in the starburst galaxies which show a highly turbulent medium compared to the spiral simulations where the density distribution is smoother. In addition, these density maps reveal an anti-correlation between the growth of instabilities and the magnitude of the external potential manifested in the angular velocity $\Omega$. For the same absolute time the low-rotation disks, SD1 and GD1, appear to be more dynamically evolved, showing a more turbulent structure and a fast departure from the initial axisymmetry. 
For example, the evolution in SD1 is so violent that the stellar feedback drags gas away from the disk, which is demonstrated by the low-density tails of the high-density blobs that are being pushed outwards. Therefore, from the notorious differences between simulations, produced by different rotational profiles, we expect to see an imprint of this behavior on the SFRs. If we compare the evolutionary states at the same number of orbits (same $\torb$ at a given radius) the contrast gets even bigger; this is contradictory with the SE law which states that the SFR at an equal number of orbits should be similar.

\subsection{Rotation}
   \begin{figure*}[ht]
	\begin{center}
	\includegraphics[width=0.95\textwidth ,natwidth=1100,natheight=500]{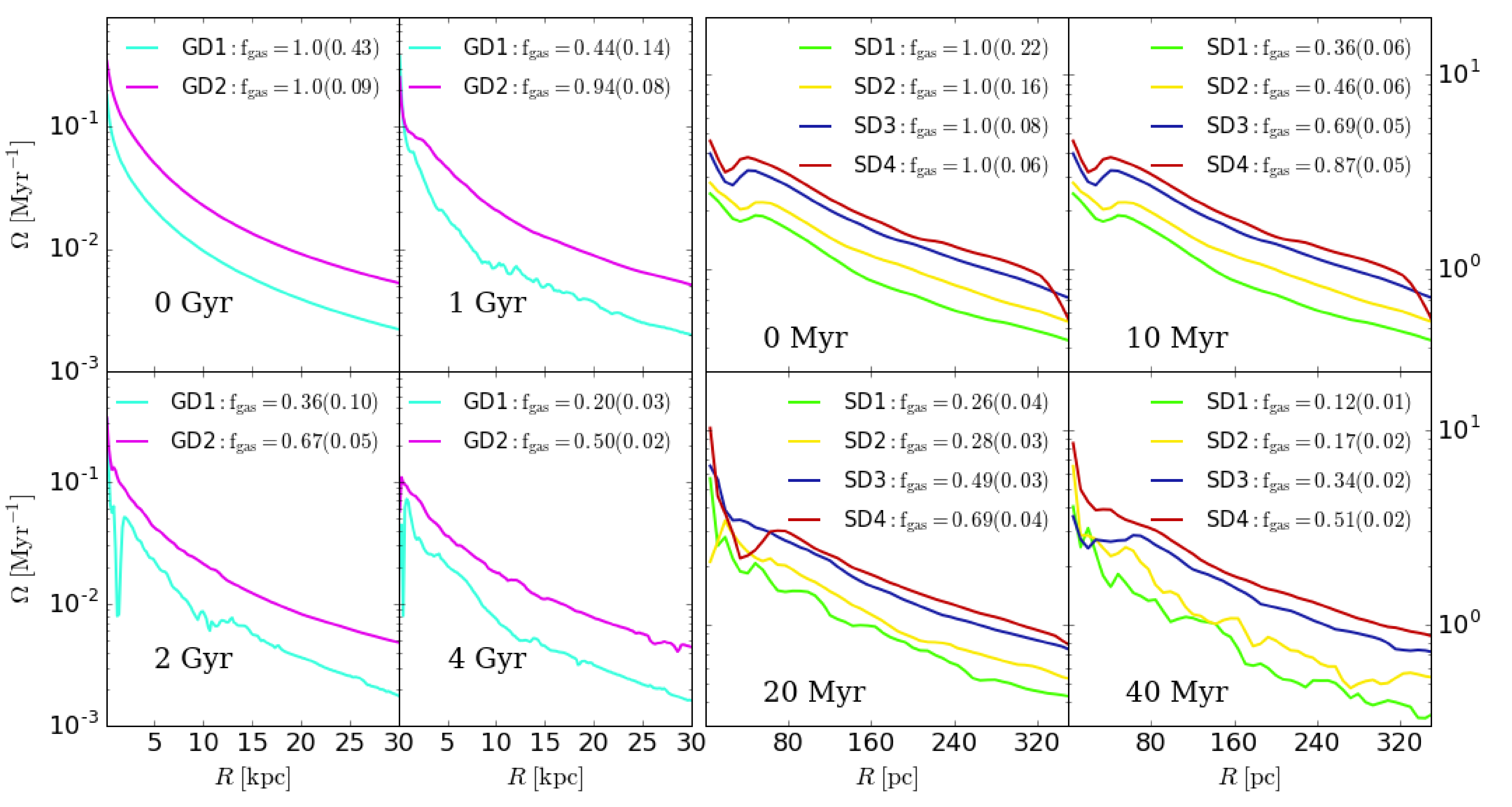}

	\caption{Radial profiles of angular velocity. Left: profiles for spiral galaxies at 0, 1, 2 and 4 Gyr. Right: profiles for starburst galaxies at 0, 10, 20 and 40 Myr. $f_{\rm gas}$ corresponds to the fraction of gas in the disk without considering the contribution from the external potentials. The quantity in brackets corresponds to the fraction of gas with respect to the total mass within a radius of 10 kpc and 300 pc for spiral and staburst galaxies respectively. For most of the time evolution the profiles of $\Omega$ remain nearly similar to their initial distribution. Significant deviations start to appear when most of the gas has been converted into stars.}
	\label{fig:Q_galaxy}
	\end{center}
	
	\end{figure*}
	
In Figure \ref{fig:Q_galaxy} we show radial profiles of the angular velocity at different times in order to visualize its variation through time. Since we want study the global behavior of the SFR as a function of angular velocity, we need to define a global value for $\Omega$. This global value has to represent the whole star-forming region. Thus it is crucial that the radial profile of $\Omega$ remains close to its initial distribution. Figure \ref{fig:Q_galaxy} also shows the gas fraction, $f_{\rm gas}$ considering only gas cells and star particles ($f_{\rm gas}=M_{\rm gas}/(M_{\rm gas}+M_{\rm stars})$). In brackets is shown the fraction of gas with respect to the total mass enclosed in a radius of 10 kpc and 300 pc for spiral and staburst galaxies respectively. The fraction $f_{\rm gas}$ is calculated in cylindrical regions with radius $R=40$ kpc and height $H=8$ kpc for spiral galaxies, and $R=500$ pc and $H= 100$ pc for starburst galaxies where $R$ is the radius and $H$ the height of the cylindrical region. We see initially that for most part of the disk, $\Omega$ has a similar functional form for each set of simulations with the main difference being the amplitude, as we intended. After the disks start to fragment and the process of star formation begins, $\Omega$ starts to deviate from its initial value. This variation is shown to be related with gas consumption due to star formation. When the gas fraction reaches values below 0.4 the disks show large variations in $\Omega$. This complicates the definition of a characteristic value of $\Omega$ which is needed for further comparison. However, at this point most of the gas has already formed stars and the remaining gas is hotter and has a lower density, taking longer time-scales to fragment. Additionally, $f_{\rm gas}$ is calculated over a wide region that includes gas that will never form stars, hence the value of $f_{\rm gas}$ in the star-forming disk is actually underestimated. In summary, for most of the evolution of these galaxies, taking a single characteristic value of $\Omega$ for a disk at a given time is a good approximation to test the effects of rotation. Additional radial profiles of $\sgas$, stellar surface density $\sstar$, $\ssfr$ and circular velocity $V_{\rm circ}$ are shown in Appendix \ref{sec:profiles}.

\section{STAR FORMATION}
\label{sec:sfl}
	\begin{figure*}
	\begin{center}
	\includegraphics[width=0.8\textwidth ,natwidth=1100,natheight=1100]{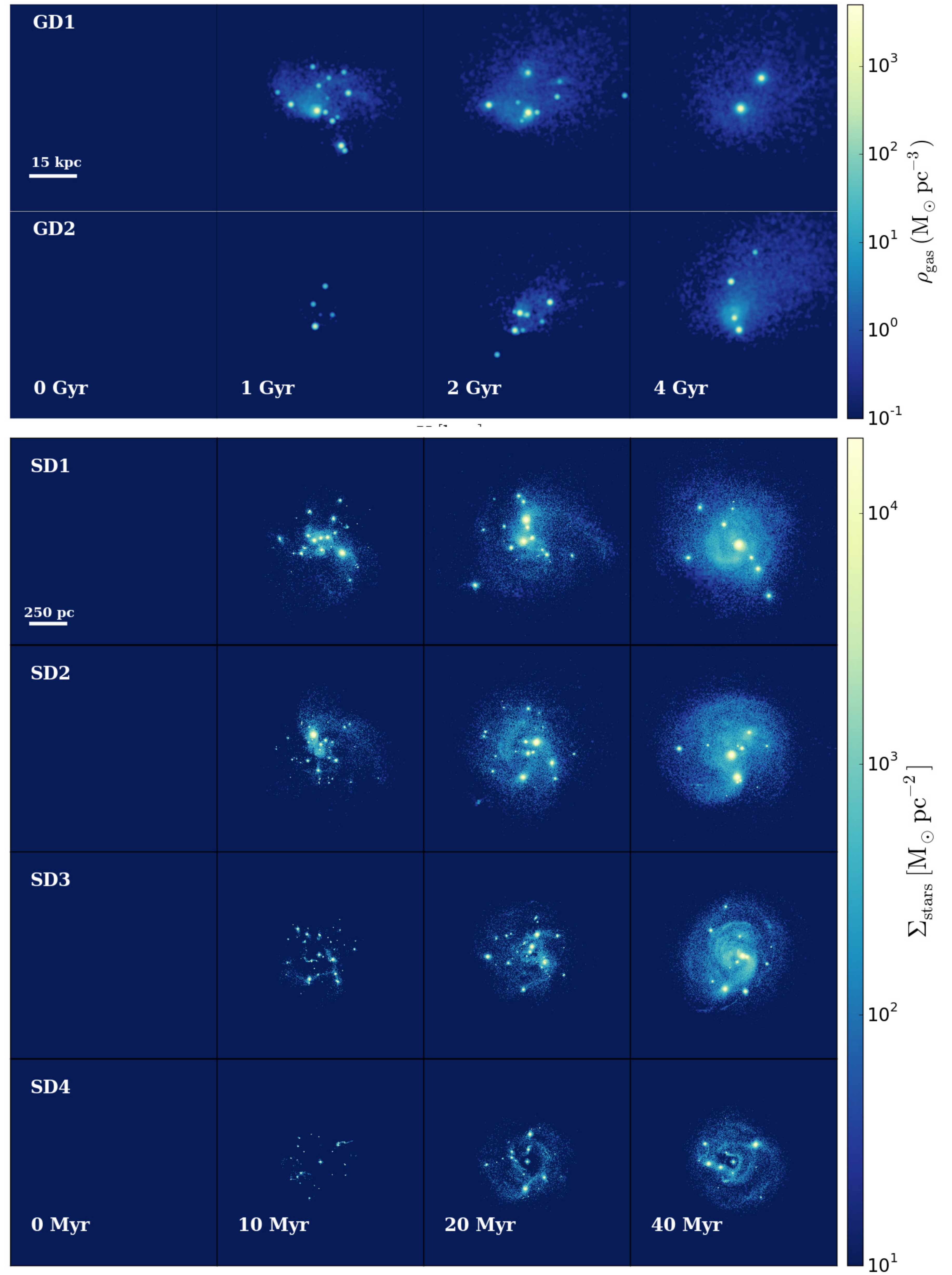}
	\caption{The panels show stellar density projections for the same models and times as in Figure \ref{fig:gas_galaxy}. Top: run GD1 have formed most of its stars at 1 Gyr while GD2 form its stars over longer time-scales. Bottom: stars are formed faster from top to bottom, which corresponds to increasing angular velocity. The size of the stellar clusters in the starburst simulations seems to be related with rotation. Both panels suggest a relation between star formation and angular velocity.}
	\label{fig:stars_nuclear}
	\end{center}
	\end{figure*}

Now we focus on the evolution of star formation and its empirical global laws. We show in Figure \ref{fig:stars_nuclear} the projected distribution of stars in the $x-y$ plane for each simulation at the same times shown in Figure \ref{fig:gas_galaxy}. In all cases we see that a high fraction of stars are located in large clusters which will disturb their initial axisymmetry. The size and mass of these star clusters are shown to be anti-correlated with the angular velocity, which is expected since the maximum unstable mass is given by $M_{\rm unstable}^{\rm max}=4\pi^4G^2 \Sigma^3 /\kappa ^4$ \citep{Escala_08}. It also evident by looking at the second snapshot of every simulation that stars are formed faster in disks with smaller rotational velocities. 

   \begin{figure*}[!ht]
	\begin{center}
	\includegraphics[width=0.95\textwidth]{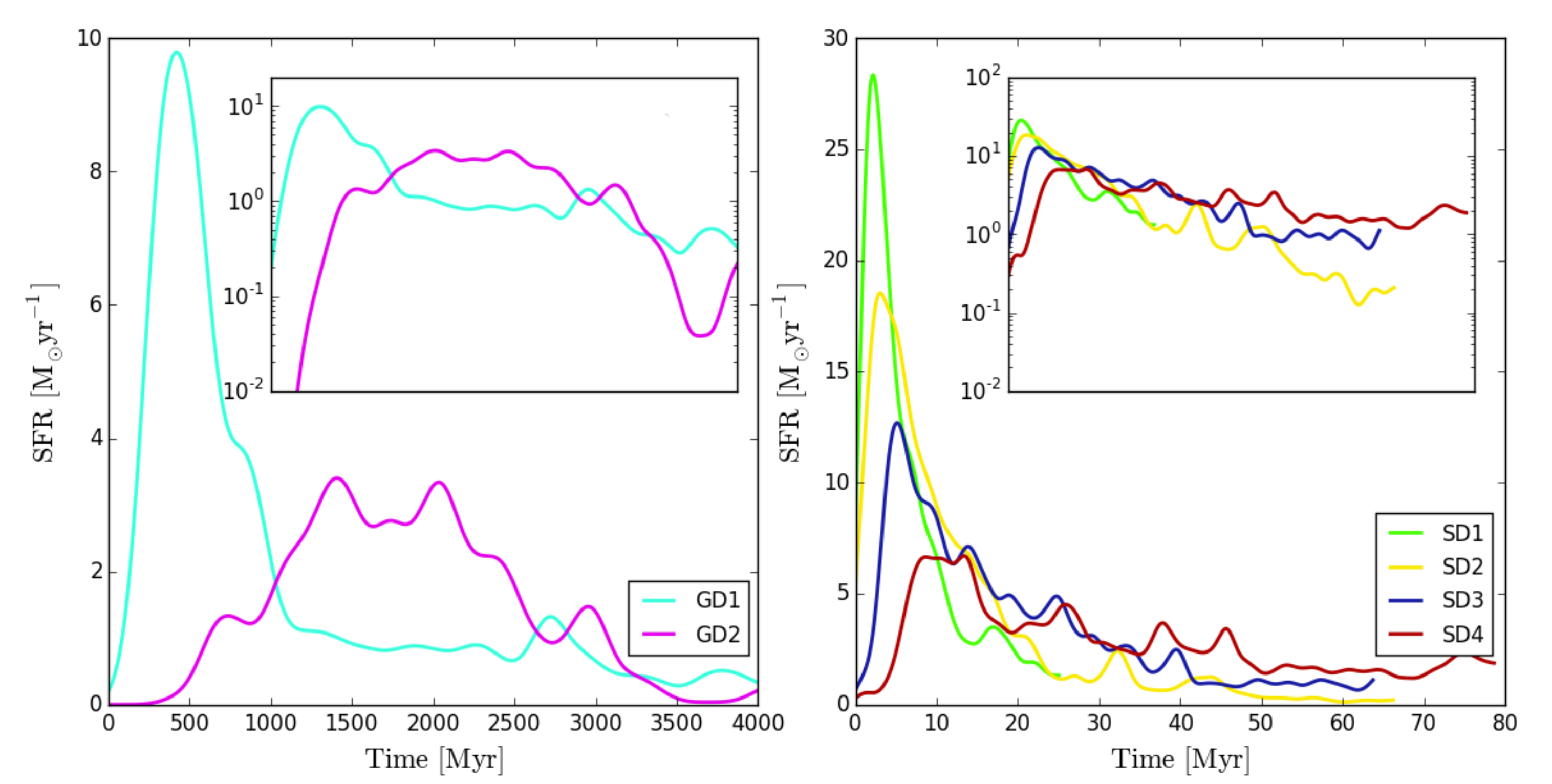}
	\caption{Time evolution of SFR in linear (main panels) and logarithmic scales (subpanels). Left: SFR for spiral galaxies as a function of time. Cyan lines corresponds to GD1 and magenta lines to GD2. GD1 reaches higher SFR rates than GD2 and at earlier times. Right: SFR for starburst galaxies against time where the green, yellow, blue, and red lines correspond to SD1, SD2, SD3, and SD4 respectively. In both sets, galaxies with lower angular velocities reach higher SFR and at earlier times.}
	\label{fig:sfrtime-galaxy}
	\end{center}
	\end{figure*}

The time evolution of star formation is shown in Figure \ref{fig:sfrtime-galaxy}. Both spiral and starburst galaxies show the same behavior: stars are formed faster for simulations with lower rotation. In addition, simulations SD1 and GD1, which correspond to the galaxies with the lowest angular velocities within their groups, reach higher SFRs than their counterparts. At late times the SFR is higher for highly rotating galaxies due to their slow gas depletion i.e. they still are able to form stars. We notice that the initial evolution of star formation is nonlinear and leads to an exponential increment of the SFR until a maximum is reached at time $t_0$. After this maximum, the SFR starts to decline exponentially as a function of time. In addition, the time to reach this exponential evolution depends strongly on the amount of rotation of the galaxy. This shows that the connection between the amount of rotation and the speed at which instabilities grows in disks \citep{Toomre_64} is also connected with the process of star formation. As we shall see in the next section there is a clear connection between the star formation efficiency and the rotation of galaxies.

\subsection{KS and SE Laws}
\label{sec:KS and SE laws}
From now on we only analyze the evolution of star formation in the exponentially decaying state. In order to identify this steady evolution of the star formation, we calculate the SFR over time convolved by a Gaussian with dispersion 100 Myr for spiral galaxies and 1 Myr for starbursts. Then we proceed to locate the time $t_0$ at which the global maximum of the SFR is reached. The time difference in the dispersion is due to the differences for both sets of galaxies in time resolution and their depletion times, which are nearly two orders of magnitude different. Finally, we only study the simulations at times $t>t_0$.\\
To compare our results with observations we have to calculate the SFR and the gas content over a projected surface, to find a law of the form 
\begin{equation}\label{eq:KSlaw}
\ssfr \propto \sgas ^N.
\end{equation}
The KS law \citep{Kennicutt_98} found empirically an index of $N\sim 1.4$, which has been interpreted as a relation between the reservoir of gas and the local free-fall time in the form:
\begin{equation}\label{eq:tfflaw}
\rho_{\rm SFR} \propto \frac{\rhogas}{\tff} \propto \frac{\rhogas}{(G\rhogas)^{-\frac{1}{2}}} \propto \rhogas^{1.5}
\end{equation}

To compute the projected quantities for the star formation laws, we need to define the integration region. We choose a cylinder whose radius encircles 90\% (hereafter $\rcyl$) of the mass in star particles created within the characteristic time of star formation $t_{\rm SFR}=M_{\rm gas}/{\rm SFR}$, where $M_{\rm gas}$ and $\rm SFR$ are the total quantities of gas and SFR for a simulation at a given time. Then in the same way we calculate the height $H_{\rm 90}$ that encloses 90\% of the stars ($H=2 \times \zcyl$). This choice for the region is for two reasons: (i) we expect that $R_{90}$ scales in a similar way as a radius based on the stellar luminosity profile and (ii) this region considers mostly the gas that will form stars. A larger radius would increase the gas reservoir with diffuse gas and reduce the efficiency spuriously, while a smaller radius would increase the uncertainties and would be affected by radial flows of gas and stars. Additionally, in order to minimize the errors introduced by changes of the integration regions, we choose the center of the simulations as the center of the cylinders. By keeping the center of the region fixed we ensure that quantities as $\Omega$ and $\sgas$ do not change drastically from one snapshot to the next. At later times, when the SFR has decreased by a factor of a few, this variation might be caused by some regions changing from passive to active within a few snapshots due to the fact that the distribution of the newly formed stars will be far from axisymmetry.

\begin{deluxetable}{cccc}
\tablecaption{Kennicutt-Schmidt Law parameters\label{table:KSlawtable}}
\tabletypesize{\scriptsize}
\tablewidth{0pt}
\tablehead{
\colhead{Run} & \colhead{Index $N$} & \colhead{Zero Point} & \colhead{Scatter} \\
\colhead{Name} & \colhead{$\Sigma_{\rm SFR} \propto \Sigma_{\rm gas} ^{\rm N}$} & \colhead{$ \Sigma_{\rm SFR} /\Sigma_{\rm gas} ^{1.4}$} & \colhead{$\Sigma_{\rm SFR} \propto \Sigma_{\rm gas} ^{1.4}$}\\
\colhead{Face-on}& \colhead{} & \colhead{(dex)} & \colhead{(dex)}}
\startdata 
SD1 & 1.996 $\pm$ 0.076 &  -2.283 $\pm$ 0.181 & 0.181 \\ 
SD2 & 1.819 $\pm$ 0.054 &  -2.434 $\pm$ 0.200 & 0.201 \\ 
SD3 & 1.947 $\pm$ 0.096 &  -2.684 $\pm$ 0.120 & 0.155 \\ 
SD4 & 1.516 $\pm$ 0.088 &  -2.835 $\pm$ 0.156 & 0.147 \\ 
GD1 & 1.807 $\pm$ 0.094 &  -3.448 $\pm$ 0.232 & 0.233 \\ 
GD2 & 2.990 $\pm$ 0.714 &  -3.964 $\pm$ 0.616 & 0.196 \\ 
\enddata
\tablecomments{Columns show the index of the best fit in the KS plane, zero point ($\ssfr$ in logarithmic scale for $\rm\sgas=1 \msun /pc^2$) and scatter with respect to a KS law with an index of $N=1.4$. The projection is carried along the $z$-axis. Disks with high angular velocity show lower zero points i.e. longer depletion times at $\rm\sgas=1 \msun /pc^2$.}
\end{deluxetable}

Within the cylinders, the SFR is computed by $\Delta m_{\star} /\Delta t$, where $\Delta m_{\star}$ is the mass in stars created within $\Delta t$ which are set to roughly 0.1$t_{\rm SF}$ (100 Myr for spirals and 1 Myr for starburst disks). For the computation of the angular velocity parameter $\Omega$ in the SE law, we measured $\Omega$ at the radius $\rcyl$. We consider this value of $\Omega$ since it is related with the radius at which the disk becomes stable against self-gravity, hence it is related with gravitational collapse and star formation.

Another restriction worth considering when calculating the SFR is the regime where SFR quickly drops due to resolution.
The distribution of the high-density cells is given by the average gas density, self-gravity, numerical resolution, and star formation through the numerical density threshold $n _{\rm thres}$ for the star formation recipe. There is a critical density where the rate at which the $n>n _{\rm thres}$ criteria is fulfilled is given by resolution. This time-scale is larger than the free-fall time of gas cells and hence the SFR declines abruptly, giving rise to a steeper power law. Then, there is a critical surface density $\Sigma_{\rm crit}$ below which the SFR is not physically correct.

In order to remove the data corresponding to $\Sigma_{\rm gas} < \Sigma_{\rm crit}$ from our analysis, we perform an orthogonal distance regression with a piecewise linear function defined by

\begin{equation}
\Sigma_{\rm SFR} = \left\{
\begin{array}{cc}
\Sigma_{\rm gas} ^{N}  & \text{if }\Sigma_{\rm gas}<\Sigma_{\rm crit}\\
\Sigma_{\rm gas} ^{1.5} & \text{if }\Sigma_{\rm gas}\geq\Sigma_{\rm crit}
\end{array}\right.
\end{equation}

 on the KS logarithmic-plane ($\Sigma_{\rm gas}$ , $\Sigma_{\rm SFR}$).  We test this function against a unique power law and we perform an $F$-test with $\alpha=0.05$; if the null hypothesis is not rejected we consider all the data from a simulation.

	\begin{figure}
     \begin{center}
	\includegraphics[width=0.45\textwidth]{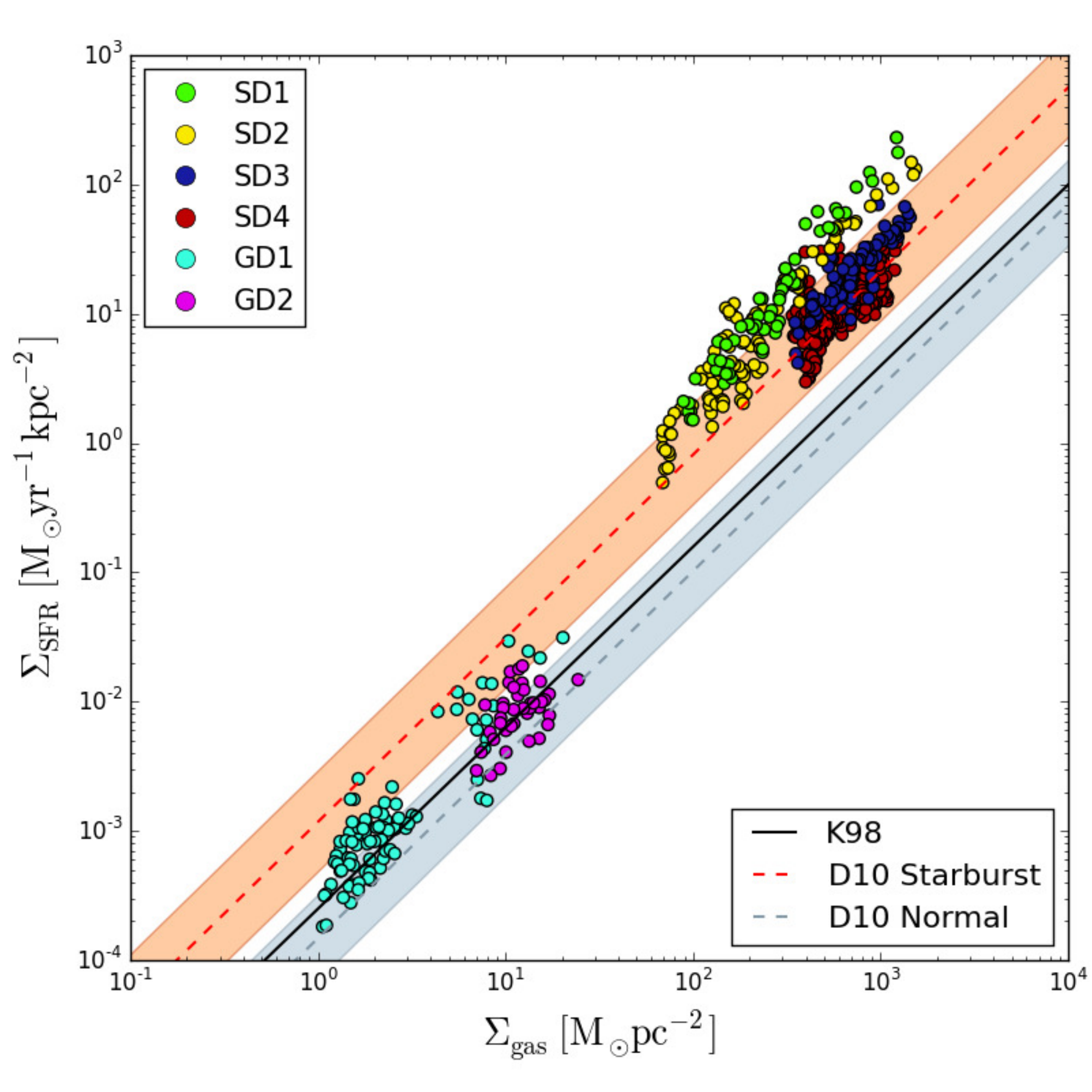}
	\caption{$\ssfr$ vs $\sgas$ for each simulation using face-on projections. Green, yellow, blue, and red circles correspond to starbursts while cyan and magenta circles represent the spiral galaxies. The black solid line corresponds to \citet{Kennicutt_98} fit while the dashed gray and red lines correspond to the \citet{Daddi_10} fits for normal and starburst galaxies respectively, with the shaded regions representing their corresponding scatter. Both sets of simulations lie close to their respective regime. Starburst galaxies show a clear trend between their depletion times and angular velocity. On average, for a given value of $\sgas$ galaxies with higher angular velocities show lower $\ssfr$.}
	\label{fig:KSlaw-face}
	\end{center}
	\end{figure}

Figure \ref{fig:KSlaw-face} shows the KS plot for our simulations during their exponentially decaying phase of star formation. For comparison with observations we have drawn the fit to the observations by \citet{Kennicutt_98} in black and the \citet{Daddi_10} fits for normal and starburst galaxies in gray and red colors respectively. A summary of the power-law index $N$, the zero point that corresponds to $\ssfr$ at 1$\rm \msun pc^{-2}$, and the scatter for our fits in Figure \ref{fig:KSlaw-face} is shown in Table \ref{table:KSlawtable}. Without considering the scatter, for most simulations the individual evolution of the SFR is well described by a KS law, with an index $N$ between $1.5$ and $2.0$. Simulation GD2 shows the largest index of the sample which might be due to the small range in $\Sigma_{\rm gas}$. 
We see a nearly parallel evolution curves for all the simulations in the ($\sgas$,$\ssfr$) plane, which is remarkable taking into account the different and complex structures shown in Figure \ref{fig:gas_galaxy}. Furthermore, our simulations display a bimodal behavior in the evolution of the surface SFR, in agreement with observations from \citet{Daddi_10}. Spiral disks show SFRs consistent with normal galaxies while starbursts show higher SFRs, more compatible with the regime of the same name.\\

The most striking feature shown in Figure \ref{fig:KSlaw-face}, and quantified by the zero point in Table \ref{table:KSlawtable}, is the anti-correlation between the depletion times and the amount of rotational support of the disk: disks with higher angular velocities show lower SFR. From a stability point of view this is expected; for high values of $\kappa$ ($\sim \Omega$) the Toomre analysis \citep{Toomre_64} tells us that the disk will be more stable. When axisymmetric systems are perturbed these perturbations will locally rotate with the epicyclic frequency and then the value of $\kappa$ can be understood as an estimation of the local rotational support. Consequently, it is desirable to analyze how rotation affects star formation.\\

{\begin{deluxetable}{cccc}
\tablecaption{Silk-Elmegreen Law parameters\label{table:silklawtable}}
\tablewidth{0pt}
\tabletypesize{\scriptsize}
\tablehead{
\colhead{Run} & \colhead{Index $N$} & \colhead{Zero Point} & \colhead{Scatter} \\
\colhead{Name} & \colhead{$\Sigma_{\rm SFR} \propto (\Sigma_{\rm gas} \Omega)^{\rm N}$} & \colhead{$\Sigma_{\rm SFR} / \Sigma_{\rm gas} \Omega$} & \colhead{$\Sigma_{\rm SFR} \propto \Sigma_{\rm gas} \Omega$}\\
\colhead{Face-on}& \colhead{} & \colhead{(dex)} & \colhead{(dex)}}
\startdata 
SD1 & 1.431 $\pm$ 0.065 &  -1.055 $\pm$ 0.214 & 0.216 \\
SD2 & 1.460 $\pm$ 0.046 &  -1.320 $\pm$ 0.252 & 0.253 \\
SD3 & 1.284 $\pm$ 0.074 &  -1.562 $\pm$ 0.129 & 0.130 \\
SD4 & 0.909 $\pm$ 0.060 &  -1.846 $\pm$ 0.139 & 0.133 \\
GD1 & 1.240 $\pm$ 0.080 &  -1.135 $\pm$ 0.271 & 0.273 \\
GD2 & 0.800 $\pm$ 0.180 &  -1.529 $\pm$ 0.172 & 0.174 \\
\enddata
\tablecomments{Columns show the index of the best fit in the SE plane, zero point ($\ssfr$ in logarithmic scale for $\rm\sgas\Omega=1 \msun /kpc^2 yr$) and scatter with respect to a SE law with an index of $N=$ 1.0. The projection is carried along the $z$-axis. Disks with high angular velocity show lower zero points i.e. smaller efficiencies.}
\end{deluxetable}}

For that we compute the most studied equation that relates $\ssfr$, $\sgas$ and $\Omega$ which is the SE law \citep{Silk_97,Elmegreen_97} 
\begin{equation}\label{eq:silk}
\ssfr \propto \sgas \Omega
\end{equation}
This relation was tested in observations by \citet{Kennicutt_98} taking $\Omega$ as the angular velocity at the optical radius.

	\begin{figure}[!htbp]
	\begin{center}
	\includegraphics[width=0.45\textwidth]{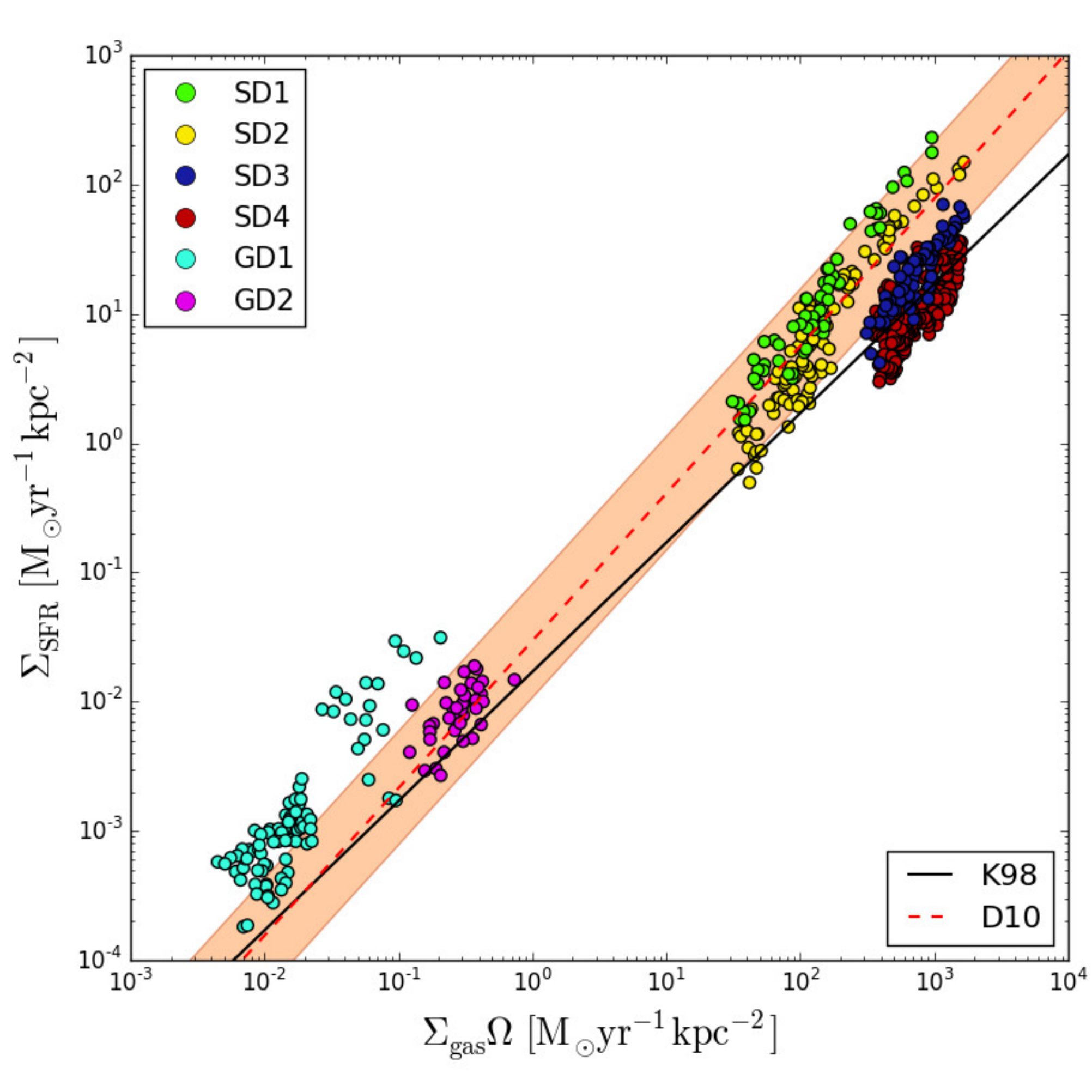}
	\caption{$\ssfr$ vs $\sgas / \Omega$ for each simulation using face-on projections.  Colors represent galaxies as in Figure \ref{fig:KSlaw-face}. The black line corresponds to \citet{Kennicutt_98} fit while the red line corresponds to the \citet{Daddi_10} fit for normal and starburst galaxies, with the shaded region representing its corresponding scatter. There is a correlation between their depletion times and angular velocity which contradicts the SE law. On average, for a given value of $\sgas/\Omega$ galaxies with higher angular velocities show lower $\ssfr$.}
	\label{fig:silklaw-face}
	\end{center}
	\end{figure}

Figure \ref{fig:silklaw-face} shows $\ssfr$ against $\sgas \Omega$ for our simulations. 
The \citet{Kennicutt_98} fit is represented by the black solid line and the \citet{Daddi_10} fit by the red dashed line. 

Table \ref{table:silklawtable} shows the parameters of the power-law function that best fits the data including the index N, the value of SFR at which $\rm\sgas=1 \msun /kpc^2 yr$ (zero point), and the scatter with respect to the SE law of index 1. The global behavior of our suite of simulations appears to follow the  SE law. However, equation \ref{eq:silk} suggests that for a constant value of $\sgas$ objects with high angular velocities would form stars more rapidly, something that is not seen in Figure \ref{fig:silklaw-face}. Although it is evident that the star formation is a function of $\Omega$ and that each individual simulation may be described by a SE law, the behavior among different simulations goes against the direct linear relation between SFR and angular velocity in disks, as suggested by the law itself. Instead, there is a consistent anti-correlation between the efficiency of the SE relation and the rotational speed; toward higher angular velocity the gas depletion time becomes larger. Even though the SE relation includes the rotational speed as an effective parameter, our simulations evince that the actual effect of rotation is not encapsulated in it. Then the SE relation is still incomplete and equation \eqref{eq:silk} works just as an estimation of the SFR for different types of galaxies but does not necessarily represent their evolution. This demonstrates that the star formation law must be a more complex function of $\Omega$ than the SE relation.

The major difference with Figure \ref{fig:KSlaw-face} is that there is a unique regime for spiral and starburst galaxies, which is easily seen in Table \ref{table:silklawtable} where the zero points of the SE law for galaxies lie within the values for starburst. This behavior is also shown in observations \citep{Daddi_10}. The slope of each fit is slightly steeper than the SE relation, with the exception of SD4 and GD2 which have a small sample of points and hence the lower index might be produced by their scatter. 

\subsubsection{The integration problem}
\label{sec:integration problem}

{\begin{deluxetable}{cccc}
\tablecaption{KS and SE law parameters (Edge-on)\label{table:KSlawtableall}}
\tablewidth{0pt}
\tabletypesize{\scriptsize}
\tablehead{
\colhead{Run} & \colhead{Index $N$} & \colhead{Zero Point} & \colhead{Scatter}}
\startdata     
& $\Sigma_{\rm SFR} \propto \Sigma_{\rm gas} ^{\rm N}$ & $ \Sigma_{\rm SFR} /\Sigma_{\rm gas} ^{1.4}$ & $\Sigma_{\rm SFR} \propto \Sigma_{\rm gas} ^{1.4}$  \\
\hline
KS-Law & & dex & dex\\
\hline                                
SD1 & 1.448 $\pm$ 0.032 &  -2.685 $\pm$ 0.122 & 0.122 \\ 
SD2 & 1.514 $\pm$ 0.036 &  -2.855 $\pm$ 0.183 & 0.183 \\ 
SD3 & 1.373 $\pm$ 0.044 &  -3.191 $\pm$ 0.138 & 0.160 \\ 
SD4 & 1.160 $\pm$ 0.046 &  -3.263 $\pm$ 0.179 & 0.165 \\ 
GD1 & 1.541 $\pm$ 0.055 &  -3.892 $\pm$ 0.210 & 0.210 \\ 
GD2 & 1.502 $\pm$ 0.199 &  -4.473 $\pm$ 0.605 & 0.205 \\ 
\hline   
 & $\Sigma_{\rm SFR} \propto (\Sigma_{\rm gas} \Omega)^{\rm N}$ & $\Sigma_{\rm SFR} / \Sigma_{\rm gas} \Omega$ & $\Sigma_{\rm SFR} \propto \Sigma_{\rm gas} \Omega$\\
\hline
SE-law & & dex & dex\\
\hline                                 
SD1 & 1.251 $\pm$ 0.032 &  -1.055 $\pm$ 0.214 & 0.216 \\
SD2 & 1.337 $\pm$ 0.031 &  -1.320 $\pm$ 0.252 & 0.253 \\
SD3 & 1.143 $\pm$ 0.037 &  -1.562 $\pm$ 0.129 & 0.130 \\
SD4 & 0.953 $\pm$ 0.041 &  -1.846 $\pm$ 0.139 & 0.133 \\
GD1 & 1.349 $\pm$ 0.054 &  -1.135 $\pm$ 0.271 & 0.273 \\
GD2 & 0.988 $\pm$ 0.137 &  -1.529 $\pm$ 0.172 & 0.174 \\
\enddata
\tablecomments{KS and SE laws parameters based on edge on integrations.  Top: Columns show the index of the best fit in the KS plane, zero point and scatter with respect to the KS law with an index of 1.4. Bottom: columns show the index of the best fit in the SE plane, zero point and scatter with respect to the SE law with an index of 1.0.}
\end{deluxetable}}

So far, the analyzed SF laws involve quantities projected on a surface area, but the physics that work in the process of star formation act in a 3-dimensional space. Hence, an arising problem with the KS and SE laws is the inclination of the disks, $i$, with respect to the line of sight, i.e. the integration axis. In the case of the KS law, if the depletion time is related to the free-fall time it assumes that there is a relation between quantities in three dimensions ($\rho _{\rm SFR}$ and $\rho _{\rm gas}$), which is preserved after the projection on 2-D space \citep{Kennicutt_98}. Just by integrating along an arbitrary axis we lose information about the compactness of the object, which is directly related to the strength of the gravitational potential that triggers the collapse. Then we expect to find objects with similar densities and SFR but with projected quantities that do not lie in the same area in the KS plane.

\begin{figure*}
	\begin{center}
	\includegraphics[width=0.9\textwidth]{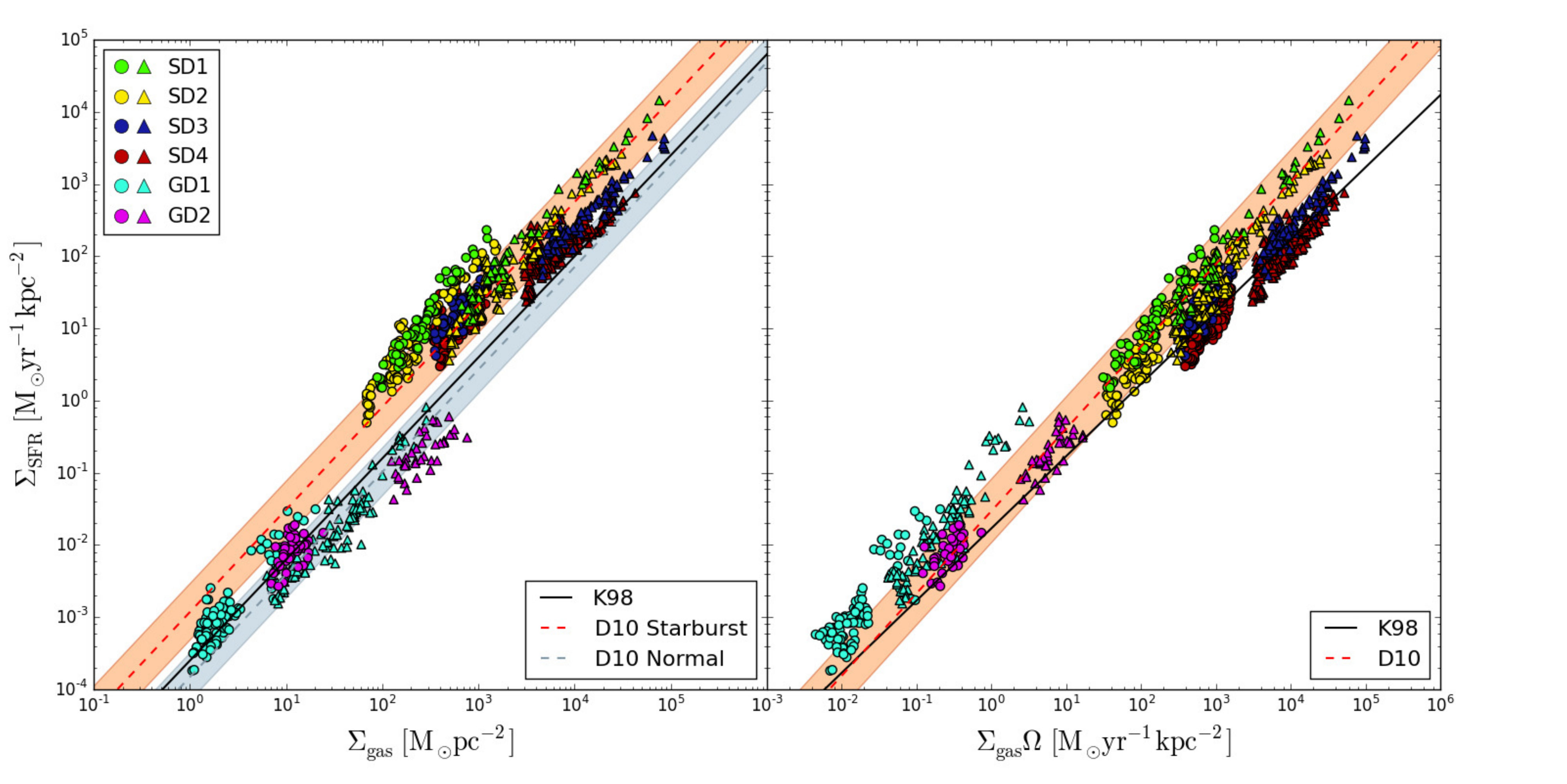}
	\caption{KS (left) and SE (right) laws for each simulation based on face-on (along the z-axis, triangles) and edge-on (along
the x-axis, circles) integrations. Colors represent galaxies as in Figure \ref{fig:KSlaw-face}. In the left panel, lines and shaded regions represent the same quantities as in Figure \ref{fig:KSlaw-face}, while in the right panel we follow the same convention as in Figure \ref{fig:silklaw-face}. For the KS law we see that different integrations produce different depletion time-scales. For the case of the SE law, different integrations produce displacements along the law.}
	\label{fig:KSlaw-all}
	\end{center}
	\end{figure*}

For this reason we perform additional edge-on integrations as an experiment to test the effects of the density distribution along the line of sight for the two extreme cases of inclination. Although in real observations edge-on galaxies are highly obscured, making it difficult to measure the total SFR and gas mass, we still think this is a question worth asking in order to test how fundamental these laws are. To compare with the same stars we choose the same cylinder volume as previously done but now the integration is carried along the $x$ axis. Figure \ref{fig:KSlaw-all} shows the KS and SE relations for face-on and edge-on configurations.

Table \ref{table:KSlawtableall} displays the best-fit parameters for the empirical relations of Figure \ref{fig:KSlaw-all}. The exponents of the KS law for edge-on integrations are more consistent with a $N=1.4$ KS law than the face-on integrations. This might be a consequence of the different projected areas: in the edge-on configuration most of the projected area is populated by new stars whereas in the face-on case there are regions devoid of star formation. In addition, the edge-on data points cover a larger range of $\Sigma_{\rm gas}$. Both effects minimize the errors in the determination of the slope of the power-law.
In the left panel of Figure \ref{fig:KSlaw-all} we see that the edge-on integrations have been shifted toward higher values of $\sgas$ and $\ssfr$ but also have been shifted toward lower efficiencies for both types of galaxies, while in the right panel the data points have been displaced along the SE law. The reason for this behavior is the same in both cases: the change is introduced by the projected surface, which is the same for both $\ssfr$ and $\sgas$. Then, by changing the integration axis, we are only multiplying both areas by the same number shifting each point along a vector of slope $1.0$. This translates in a shift below and along the empirical laws for the KS law and the SE law respectively. An important point to take from here is that similar objects of the same age but with different inclinations with respect to the line of sight lie on a line with a slope $\sim 1.0$ for both relations.

\subsection{A dimensionally homogeneous law}
\label{sec:A dimensionally homogeneous Law}

Since we are studying the SFR and its relations with galactic properties it is fundamental to know how to formulate such relations in a physically meaningful way. As a starting point, it is desirable to have some knowledge of the parameters involved in this process and how they would be related in an equation beforehand. The tentative functional forms of such relations can be found by using dimensional analysis tools. According to these, any physically meaningful equation must be dimensionally homogeneous: only commensurable quantities (quantities with the same dimensions) may be compared, equated, added, or subtracted. Therefore if we want to find a law for $\ssfr$ and its fundamental parameters ($G$, $\sgas$, $\Omega$, $R$, etc.) dimensional homogeneity needs to be fulfilled. An important theorem to find such equations is the Buckingham $\Pi$ theorem of dimensional analysis. This theorem was applied by \citet{Escala_15} to find a star formation equation expressed by:
\begin{equation}
\ssfr=\epsilon \sqrt{\frac{G}{L}} \sgas ^{3/2}
\end{equation}
(E15 as mentioned above) where $L$ represents a characteristic length scale of the system. \citet{Escala_15} tested two characteristic scales for a disk system; the radial and vertical length scales in cylindrical coordinates showing that the latter results in a unique law that represents the SFR for different types of galaxies. Furthermore, he showed that depending on the physics that set the parameter $L$, such as the characteristic size of the instabilities, it is possible to recover different star formation laws found in the literature. For instance, replacing $L$ by the largest scale not stabilized by rotation, $\lrot =4\pi^2G \Sigma /\kappa^2$ \citep{Escala_08}, results in the SE law. \citet{Escala_15} also states that the efficiency $\epsilon$ is given by physics at smaller scales and that it must be a function of dimensionless parameters. In agreement with this hypothesis, it has been found and tested in simulations that $\epsilon$ is likely to be related to the dynamics inside molecular clouds through dimensionless quantities \citep{Padoan_12,Semenov_15}.

As in the KS and SE laws, we compute the E15 relation integrating over the $z$ and $x$ axes. For each integration axis we choose $L$ as the gas mass-weighted dispersion $L=\sqrt{\langle s^2 \rangle -\langle s \rangle ^2}$, where $s$ is the position along the line of sight and $\langle f \rangle $ is defined by

\begin{equation}
\displaystyle \langle f \rangle =  \frac{\int _V \rho(\vec{x}) f dx^3}{\int _V \rho(\vec{x}) dx^3}
\end{equation}

\begin{figure}
	\begin{center}
	\includegraphics[width=0.45\textwidth]{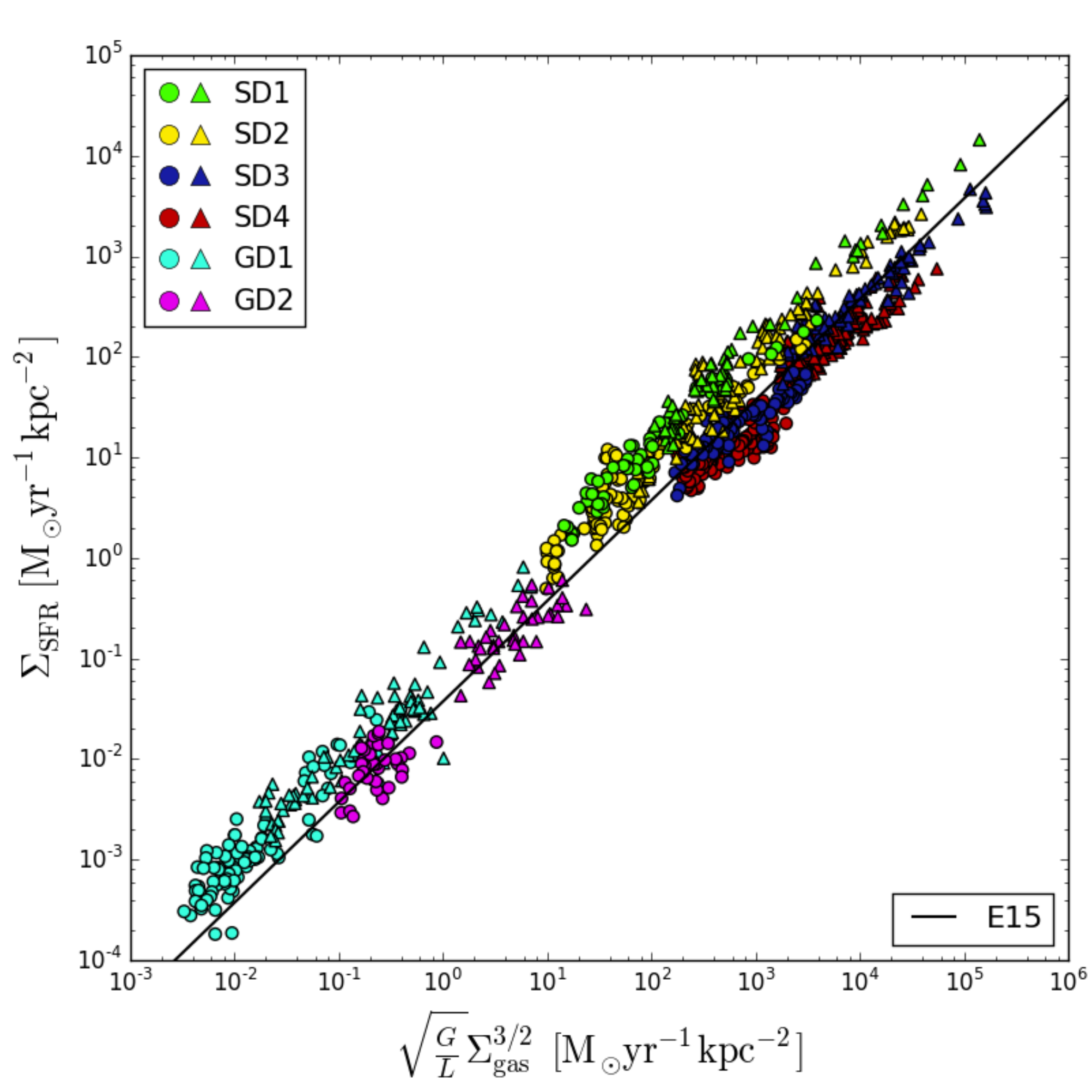}
	\caption{E15 relation for both projection along the z-axis (circles) and the x-axis (triangles). Green, yellow, blue, and red colors correspond to starburst while cyan and magenta colors represent the spiral galaxies. The black solid line shows the observational best fit for the E15 relation. Different integrations (circles and triangles) result in displacement along this relation.}
	\label{fig:escalalaw}
	\end{center}
	\end{figure}

In this expression, the volume of integration $V$ corresponds to the cylinder defined by $\rcyl$ and $\zcyl$. The results are shown in Figure \ref{fig:escalalaw}, which displays the \cite{Escala_15} equation for the whole set of simulations.  In addition, a description of the fit parameters is shown in Table \ref{table:escalatable}. \\

{\begin{deluxetable}{cccc}
\tablecaption{Escala (2015) law parameters (Edge-on)\label{table:escalatable}}
\tablewidth{0pt}
\tabletypesize{\scriptsize}
\tablehead{
\colhead{Run} & \colhead{Index $N$} & \colhead{Zero Point} & \colhead{Scatter}\\
\colhead{Name}& \colhead{$\Sigma_{\rm SFR} \propto \left(\sqrt{\frac{G}{L}} \Sigma_{\rm gas} ^{3/2}\right)^{\rm N}$} & \colhead{$\Sigma_{\rm SFR} /\sqrt{\frac{G}{L}} \Sigma_{\rm gas} ^{3/2}$} & \colhead{$\Sigma_{\rm SFR} \propto \sqrt{\frac{G}{L}} \Sigma_{\rm gas} ^{3/2}$}}
\startdata     
\hline
Face-on & & (dex) & (dex)\\
\hline                                
SD1 & 0.898 $\pm$ 0.032 &  -0.909 $\pm$ 0.143 & 0.143\\
SD2 & 0.976 $\pm$ 0.029 &  -1.053 $\pm$ 0.182 & 0.182\\
SD3 & 0.699 $\pm$ 0.039 &  -1.499 $\pm$ 0.193 & 0.178\\
SD4 & 0.692 $\pm$ 0.038 &  -1.538 $\pm$ 0.190 & 0.204\\
GD1 & 1.069 $\pm$ 0.047 &  -1.056 $\pm$ 0.202 & 0.202\\
GD2 & 1.227 $\pm$ 0.244 &  -1.436 $\pm$ 0.210 & 0.210\\
\hline
Edge-on  & & (dex) & (dex)\\
\hline                                
SD1 & 0.950 $\pm$ 0.021 &  -0.852 $\pm$ 0.129 & 0.129\\
SD2 & 0.999 $\pm$ 0.024 &  -1.012 $\pm$ 0.183 & 0.183\\
SD3 & 0.826 $\pm$ 0.026 &  -1.397 $\pm$ 0.178 & 0.159\\
SD4 & 0.812 $\pm$ 0.028 &  -1.488 $\pm$ 0.172 & 0.185\\
GD1 & 0.974 $\pm$ 0.038 &  -1.067 $\pm$ 0.233 & 0.233\\
GD2 & 0.836 $\pm$ 0.101 &  -1.466 $\pm$ 0.247 & 0.206\\
\enddata
\tablecomments{\footnotesize Columns show the index of the best fit in the E15 plane, the\\
 zero point ($\ssfr$ in logarithmic scale for $\rm\sqrt{\frac{G}{L}} \Sigma_{\rm gas} ^{3/2}$=\\$\ 1 \msun {\rm kpc^{-2} yr^{-1}}$ or $\log_{10}\epsilon $) and scatter with respect to the E15 relation.\\ Different integrations show 
  similar zero points.}
\end{deluxetable}}

Figure \ref{fig:escalalaw} shows the \cite{Escala_15} equation for the whole set of simulations, and a description of the fit parameters is shown in Table \ref{table:escalatable}. The slopes displayed in Table \ref{table:escalatable} show that the evolution of the SFR is well depicted by the E15 relation. At the same time, the slopes are more consistent compared to the KS plot, which agrees with a dimensionally homogeneous equation. Although the functional form of the E15 relation is similar to the KS relation there is a major difference: the E15 relation suppresses the bi-modal behavior shown by the KS law and demonstrates that the global behavior of the SFR is well represented by a single regime. This shows that the gas distribution along the line of sight must be considered if we want to compare objects with different three-dimensional geometries. Most remarkably, this relation shows that for both lines of sight the evolution of each galaxy follows a unique path, and hence a unique observed efficiency, suggesting that this homogeneous relation is taking into account the density distribution along the line of sight in the correct form. One interpretation is that with the inclusion of the extra parameter $L$ we are actually fitting a law of the form $\rho _{\rm SFR} \propto \frac{\rho _{\rm gas}}{\tff} \propto \sqrt{G} \rho _{\rm gas}^{1.5}$ which we can express in terms of the integrated densities along different axes like $x$ and $z$:
\begin{equation}
 \frac{\Sigma _{{\rm SFR} ,i}}{L_i} \propto \sqrt{G}\left(\frac{\Sigma _{{\rm gas},i}}{L_i} \right)^{1.5}
\end{equation}
where $i$ is the integration axis. Then if the length scale $L$ is chosen properly, the observed exponent and star formation efficiency have the same values for any observer.

However, as seen in the previous section, Figure \ref{fig:escalalaw} also shows that simulations with different angular velocities lie on parallel distinctive curves, manifesting that this formulation does not incorporate the net effect of galactic rotation in the star formation process. Knowing that the E15 relation incorporates the effects of the density distribution, we proceed to study the efficiency of star formation and its dependence with $\Omega$.

\subsection{Inclination of galaxies}
\label{sec:inclination}

	\begin{figure*}[t]
	\begin{center}
	\includegraphics[width=1.0\textwidth]{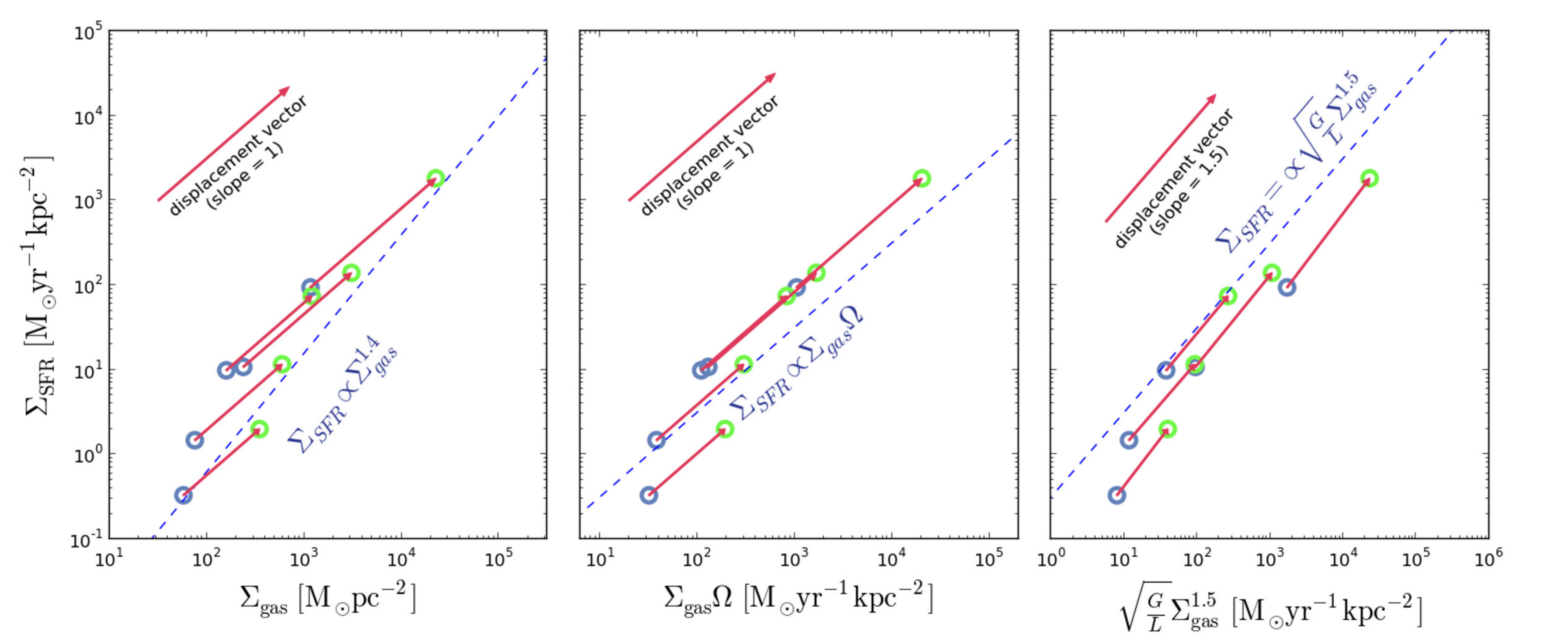}
	\caption{Displacement vectors due to effects of the line of sight. To illustrate this, we show data points from the simulation
SD2. Green and blue circles correspond to integrations along the x and z axis respectively. Left: graph showing displacements in the KS plot. The blue dashed line has a slope of 1.4. Middle: graph showing displacements in the SE plot. The blue dashed line has a slope of 1. Right: graph showing displacements in the E15 plot. The blue dashed line has a slope of 1.}
	\label{fig:disp3}
	\end{center}
	\end{figure*}

As mentioned in Section \ref{sec:integration problem} the change in the inclination of galaxies produces a vector displacement, in the KS and SE plots, which has a linear slope ($N=1$). To illustrate this Figure \ref{fig:disp3} shows the displacement for data points of SD2 in the KS, SE, and E15 plots. From this figure we see clearly that both the KS and SE laws are  good models of the SFR for a single galaxy. However, the efficiency of the KS law depends on the line of sight of the observer while the efficiency of the SE law appears to be independent of it. For the latter we emphasize that this behavior is only due to the fact that a change in the projected area is also linear. On the other hand, for the E15 relation this displacement occurs parallel to the law despite not being a linear power law. The requirement of homogeneity leads us directly to a law that is both a good model for the individual evolution of the SFR and an equation independent of the observer. The orders of magnitude for this shift, considering inclinations from $i=0^{\circ}$ to $i=90^{\circ}$, are proportional to $R/H$ where $R$ is the radius of the object and $H$ the scale height. For starburst disks $R/H$ varies from 10 to 100 while for spirals this value is always $\sim$ 10.

\subsection{Efficiency}

	\begin{figure*}[ht!]
	\begin{center}
	\includegraphics[width=1.0\textwidth ,natwidth=1100,natheight=550]{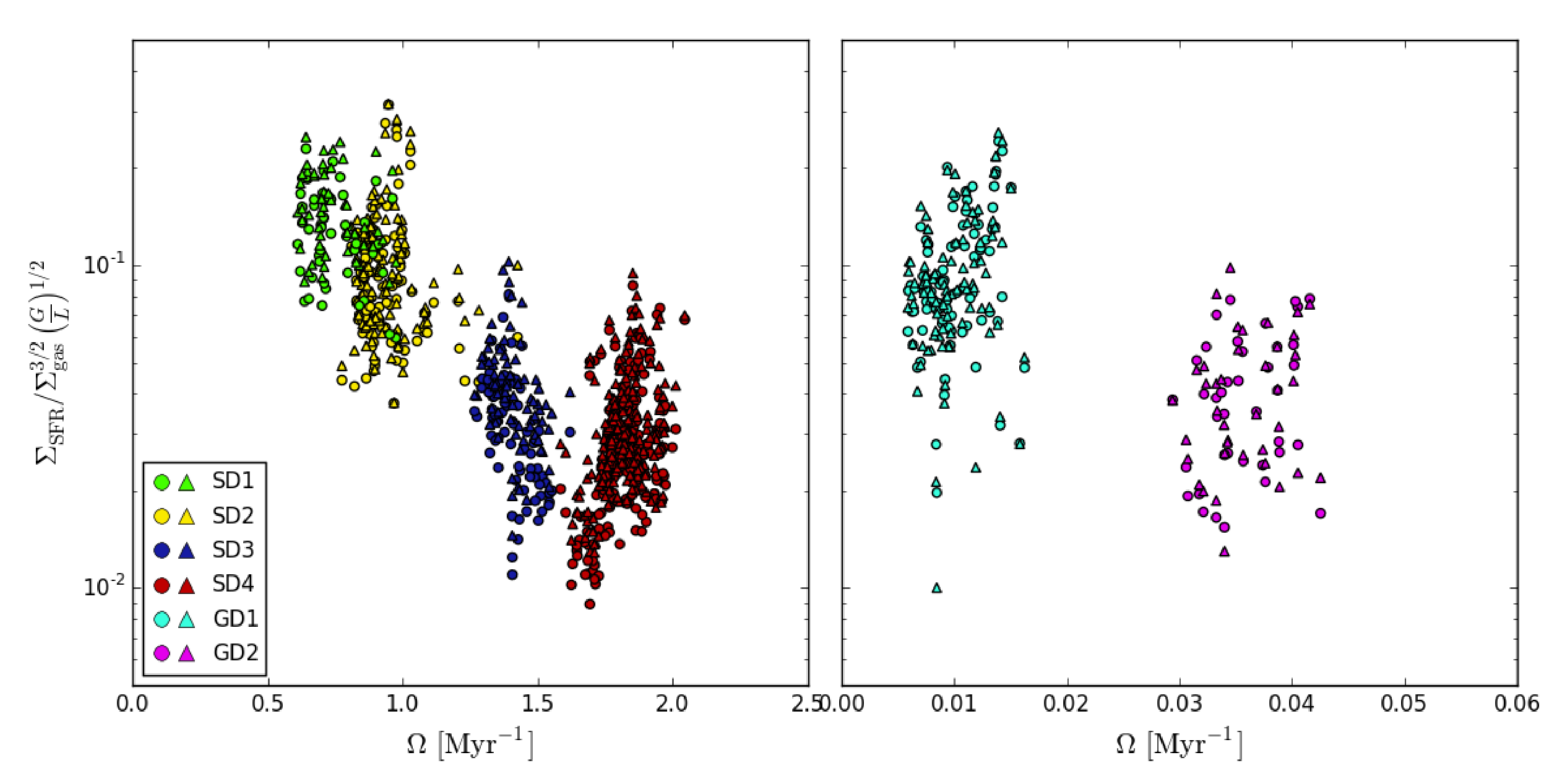}
	\caption{Efficiency vs $\Omega$: efficiency of the E15 relation against the characteristic value of $\Omega$. Circles and triangles correspond to projections along the z-axis and x-axis respectively. Left: Starburst galaxies are displayed in green, yellow, blue, and red, which correspond to SD1, SD2, SD3 and SD4, respectively. Right: spiral galaxies are shown in cyan and magenta, which correspond to GD1 and GD2, respectively. In each panel, the tar formation efficiency seems to exponentially decrease with $\Omega$.}
	\label{fig:eff}
	
	\includegraphics[width=1.0\textwidth ,natwidth=1100,natheight=550]{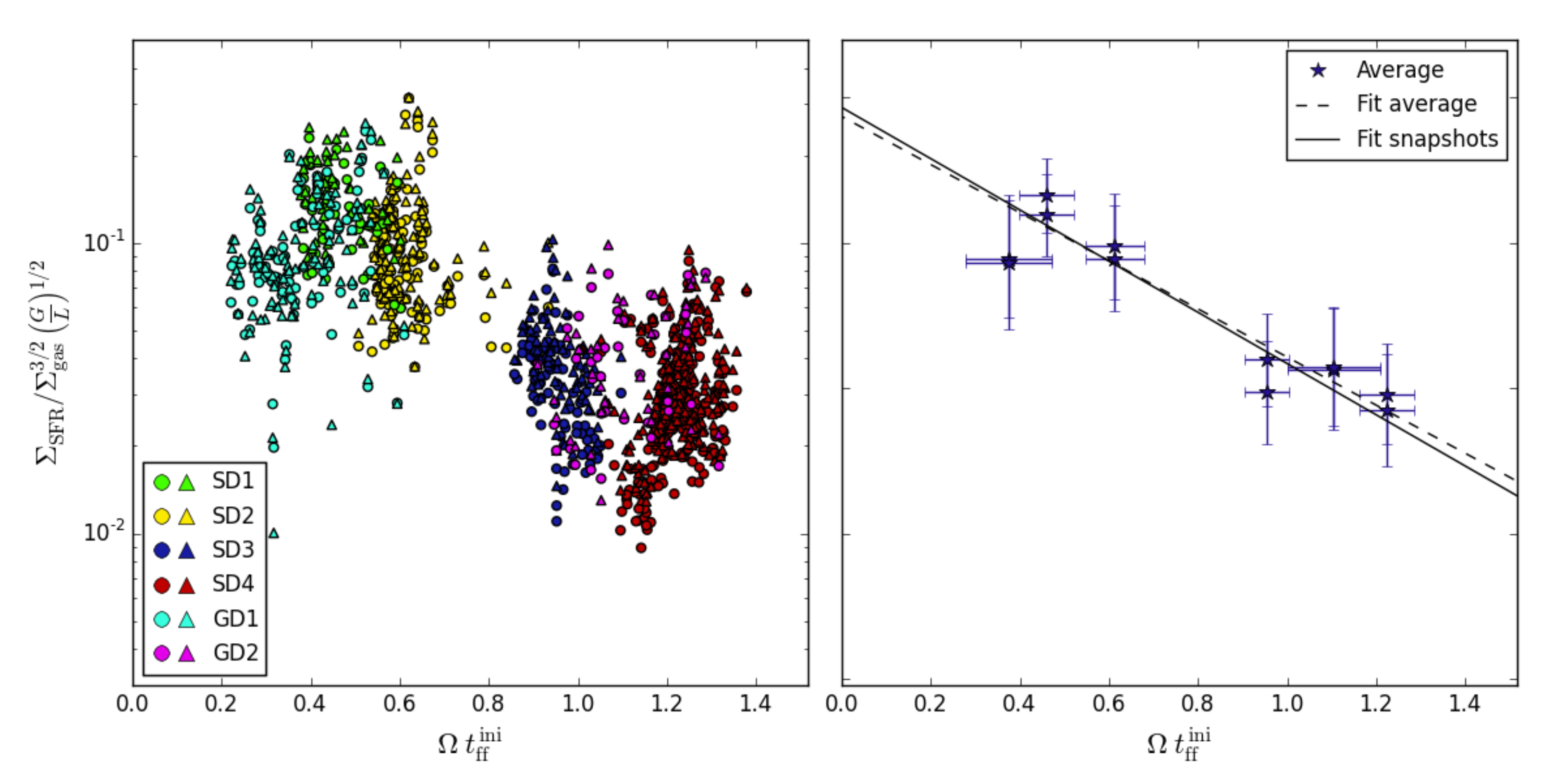}
	\caption{Efficiency vs $\Omega \tff ^{\rm ini}$ : Left: symbols are as in Figure \ref{fig:eff}. Right: stars corresponds to the mean values for each individual run and inclination. The dashed line corresponds to the fit to the mean values, which is approximately given by 0.27 exp$(-1.91\Omega \tff ^{\rm ini})$, while the solid line corresponds to the fit to all the points in the left panel, given by 0.29 exp$(-2.03\Omega \tff ^{\rm ini})$.}
	\label{fig:eff-omega}
	\end{center}
	\end{figure*}
	
	\begin{figure}
	\begin{center}
	\includegraphics[width=0.45\textwidth]{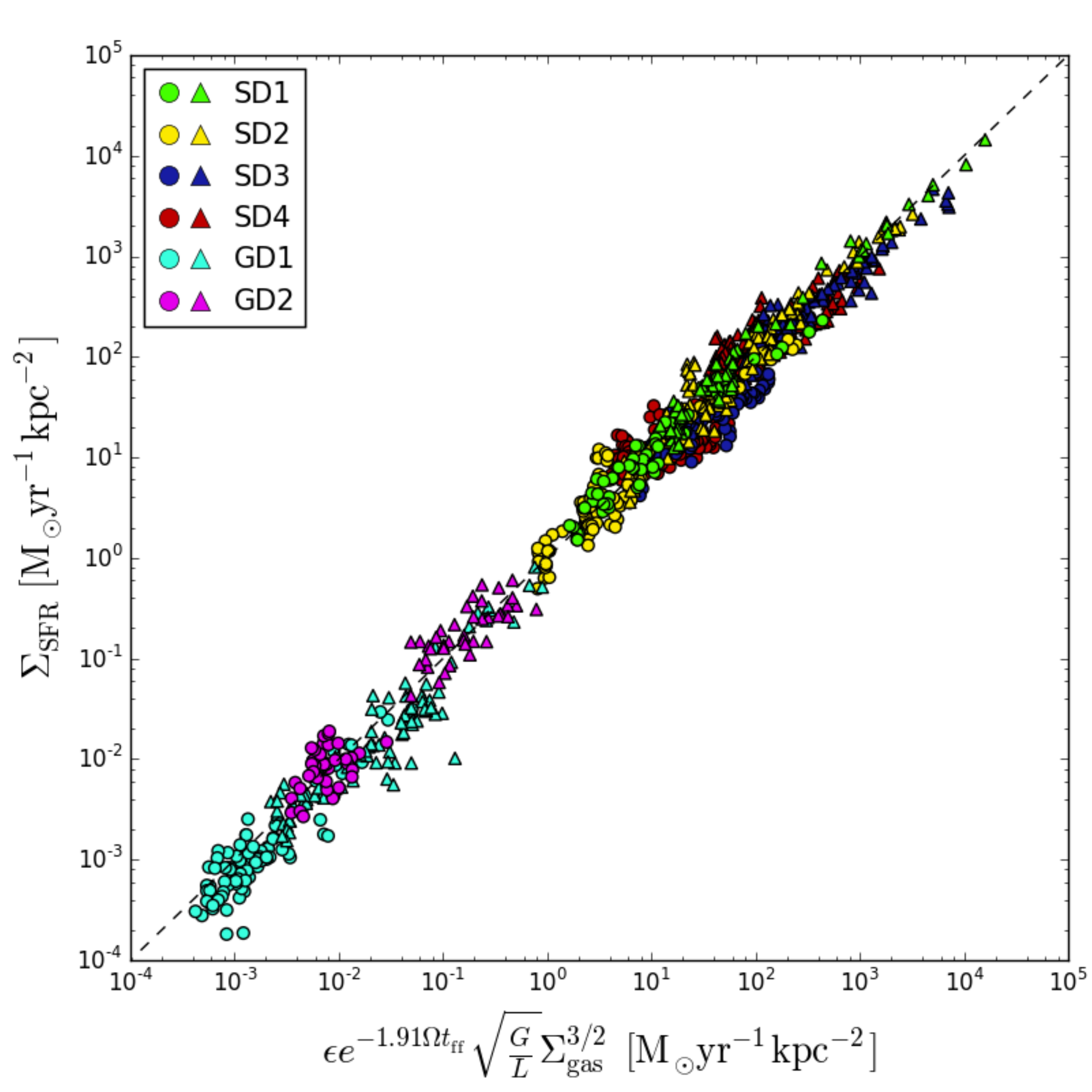}
	\caption{This work: $\ssfr$ as a function of 0.27 exp $(-1.91\Omega \tff ^{\rm ini})\sqrt{\frac{G}{L}}\sgas ^{3/2}$. Symbols are as in Figure \ref{fig:escalalaw}. The black dashed line represents the equality. The whole suite of simulations lies along a unique regime of star formation for edge-on and face-on integrations.}
	\label{fig:work}
	\end{center}
	\end{figure}

As shown in Figure \ref{fig:sfrtime-galaxy} there is a connection between the angular velocity of galaxies and the time needed to reach an equilibrium state of star formation (exponential decay). Since the first stage of evolution corresponds to the fragmentation and growth of over-densities there might be a connection between the growth-rate of instabilities and $\Omega$. Figures \ref{fig:KSlaw-all} and \ref{fig:escalalaw} also show that the amount of rotation is intrinsically connected with the star formation efficiency. This shows that galactic rotation plays an important role in the evolution of galaxies.\\
In order to construct a more robust star formation law that includes the observed effect of rotation in our simulations, we compare the efficiency of the E15 relation to the averaged angular velocity of the disks.
In this case, we choose the representative value of $\Omega$ in a disk as the one associated with the total angular momentum, which is associated with the stability of a gaseous disk \citep{Toomre_64}. We compute this angular velocity as $J_z/I_z$, where $J_z$ and $I_z$ are the $z$-components of the angular momentum and the moment of inertia respectively. We choose this parameter instead of $\Omega$ at $\rcyl$ or at a fixed radius, $R_{\rm fixed}$, since $\Omega_{\rcyl}$ varies with time and a $\Omega(R_{\rm fixed})$ has less physical meaning for a star formation region of variable size. Additionally, both $\Omega_{\rcyl}$ and $\Omega(R_{\rm fixed})$ do not consider the distribution of gas. Accordingly, we consider our choice of $\Omega$ to be more representative. We show the effects of choosing a different definitions of the characteristic $\Omega$ in the Appendix \ref{sec:Omegas}.

Figure \ref{fig:eff} shows the star formation efficiencies of the E15 relation for face-on and edge-on integrations against $\Omega$. Starburst and spiral galaxies lie on different ranges of $\Omega$ hence displaying different efficiency curves. The functional form of the efficiency can be approximated by two separated exponentially decreasing functions, one for spirals and another for starburst galaxies. In mathematical terms, the star formation relation is approximately given by $\ssfr = \exp (-B\Omega) \sqrt{G/L} \sgas^{3/2}$ , where the value of the constant $B$ depends, in principle, on whether this is a spiral or starburst disk. The requirement of dimensional homogeneity demands an efficiency that depends only on dimensionless quantities, implying the existence of another time-scale $t^* \sim B$ that should normalize this function and represent the efficiency for any kind of object. For a body that lacks rotational support, the free-fall time-scale controls the rate of gravitational collapse, thus $\tff$ is a natural candidate for $t^*$. Since the current density of the collapsing structures is directly affected by the rotational support, the simplest approach is to take the free-fall time of the simulation as done in previous works \citep[e.g.][]{Padoan_12}, as a proxy for this characteristic time-scale related to a system with no rotation. To remove the effects of our initial conditions we measure the free-fall time after the initial vertical free-fall. For this we do the following: we start by defining the average free-fall time as $\bar{t}_{\rm ff} = (G\bar{\rho})^{-1/2}$, where $\bar{\rho}$ is the volume-weighted average of the density field. While the disk starts to cool down, it will experience vertical collapse with a characteristic time $\bar{t}_{\rm ff,0}$. For times $t>\bar{t}_{\rm ff,0}$ the vertical density distribution is less affected by the initial conditions. For this reason we measure $\bar{t}_{\rm ff}$ at a time $t=\bar{t}_{\rm ff,0}$, where the disk height reaches a steady value. We refer to this time-scale as $\tff ^{\rm ini}$.\\

We plot the efficiency against $\Omega \tff ^{\rm ini}$ for all the simulations, which is shown in Figure \ref{fig:eff-omega}. The time-scale $\tff ^{\rm ini}$ does a great job rescaling the effects of rotation, which allows us to find a unique function relating the star formation efficiency and the dimensionless quantity $\Omega \tff ^{\rm ini}$. The efficiency is well represented by an exponential profile as found by \citet{Padoan_12} with $\epsilon \propto \exp(-\beta \tff/\tdyn)$ where $\tff \propto (G\rho)^{-1/2}$. We also fit an exponential curve to the data an find the functions:
\label{subsec:efficiency}

	\begin{subequations}
		\begin{equation}
		\epsilon=0.27_{-0.04}^{+0.05}e^{(-1.91\pm 0.22)\Omega \tff ^{\rm ini}}
		\label{eq:eff-aver}
		\end{equation}
		\begin{equation}
		\epsilon=0.29_{-0.01}^{+0.01}e^{(-2.03\pm 0.05)\Omega \tff ^{\rm ini}}
		\label{eq:eff-all}
		\end{equation}
	\end{subequations}

where equation \ref{eq:eff-aver} is the best fit to the averages computed for each simulation and equation \ref{eq:eff-all} is the best fit to all the points in Figure \ref{fig:eff-omega}. 




Based on this expression, we show in Figure \ref{fig:work} that the global behavior of the star formation in
our simulations can be represented by a unique relation
\begin{equation}
\label{eq:work1}
\ssfr = 0.27e^{-1.91\Omega \tff ^{\rm ini}} \sqrt{\frac{G}{L}}\sgas ^{3/2}
\end{equation}

\begin{deluxetable}{cc}
\tablecaption{Scatter of star formation relations\label{table:scatter}}
\tabletypesize{\small}
\tablewidth{0pt}
\tablehead{
\colhead{SFRRrelation} & \colhead{Scatter}\\
\colhead{Name} & \colhead{(dex)}}
\startdata 
Kennicutt-Schmidt  & 0.490\\
Bimodal KS & 0.360\\
Silk Elmegreen & 0.362\\
Escala (2015) & 0.316\\
This work & 0.206 \\
\enddata
\tablecomments{Scatter in the star formation rate for the different laws analysed in this work and our formulation.}
\end{deluxetable}

which is independent of galaxy type, line of sight, and incorporates the effect of the angular velocity. This star formation relation reduces the scatter of the classical KS law by 58\% and that of the bi-modal KS law by 42\% (see Table \ref{table:scatter}). The proposed star formation relation in equation \eqref{eq:work1} arises to the same functional form discussed in Section 2.1 of \citet{Escala_15} with an efficiency given by $\epsilon \propto \exp(-1.6 \tff/\tdyn)$. An efficiency of this kind was already proposed by \citet{Padoan_12} and tested by \citet{Semenov_15}, where the dynamical time $\tdyn$ corresponded to the turbulent crossing time $t_{\rm cr}$. In the simulations presented here, $t_{\rm cr}$ is replaced by the inverse of $\Omega$, change that is expected since we are modeling global disk dynamics and $\Omega ^{-1}$ arises as a natural time-scale. In contrast, in simulations of isolated boxes subject to supersonic turbulent forcing \citep{Padoan_12}, which model individual giant molecular clouds, $t_{\rm cr}$ is the only possible choice for $\tdyn$. In other words, we have linked global disk dynamics to the local dynamics of giant molecular clouds. Moreover, the constant factors 1.61 and 1.91 are of the same order despite coming from completely different numerical experiments. In Appendix \ref{sec:Omegas} we show that this result does not vary significantly when choosing different definitions for the characteristic angular velocity.\\

\subsection{Normalizing time-scale}
\label{sec:Norm}
	\begin{figure*}[!htbp]
	\begin{center}
	\includegraphics[width=0.9\textwidth]{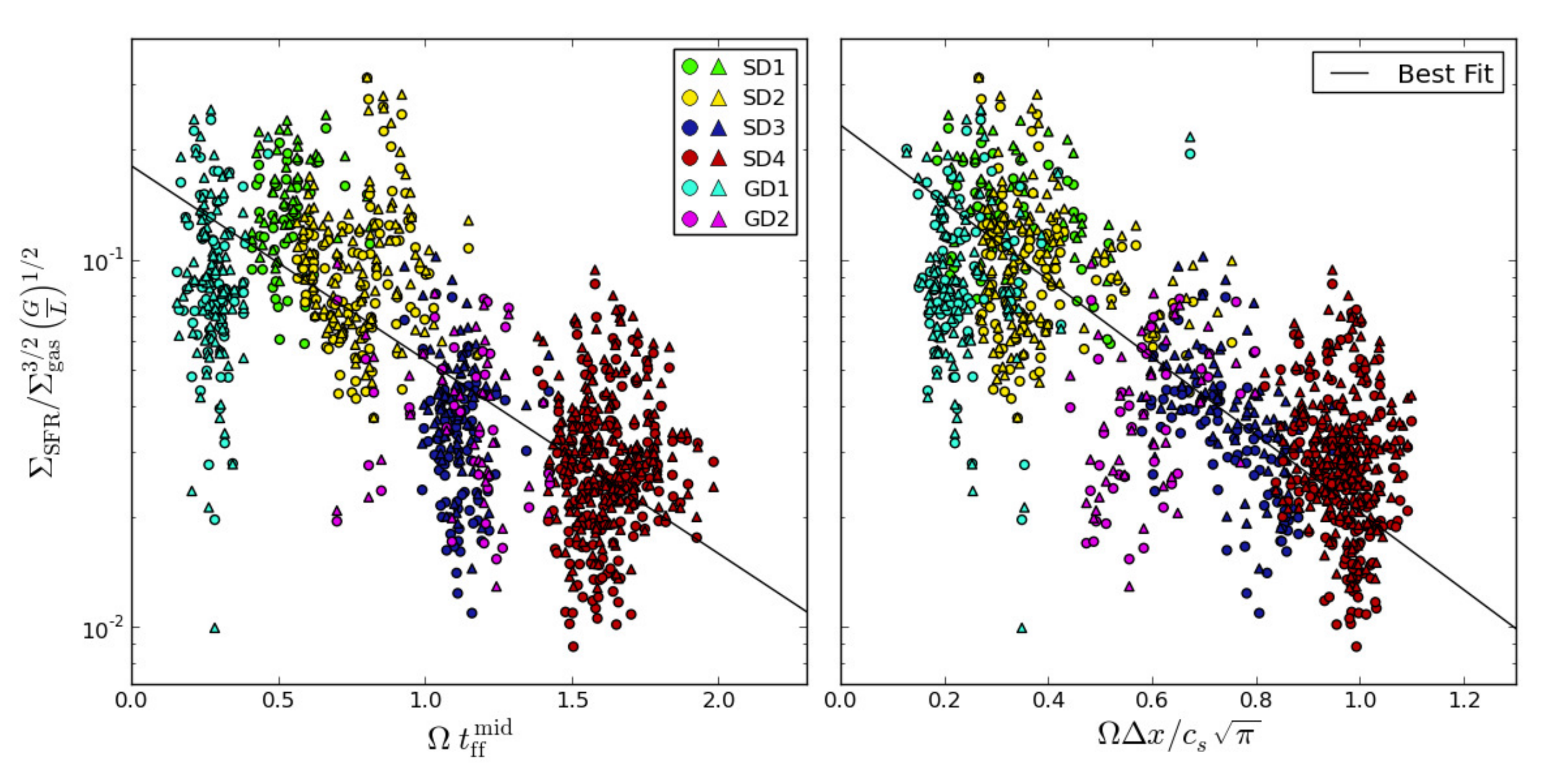}
	\caption{Alternative normalizing time-scales: {\it Left}: Plot showing the efficiency of the E15 equation as a function of $\Omega t_{\rm ff}^{\rm mid}$ where $t_{\rm ff}^{\rm mid}$ is the mass-weighted average of the free-fall time at the midplane. Triangles and circles correspond to integrations along the x and z axis respectively. The solid line represent the best fit with a slope of -1.22. {\it Right}: Plot showing the efficiency of the E15 equation as a function of $\Omega \Delta x /c_s \sqrt{\pi}$ where $\Delta x$ is the size of the smallest cell and $c_s$ is the mass-weighted average of the sound speed. The solid line represent the best fit with a slope of -2.43.}
	\label{fig:norm}
	\end{center}
	\end{figure*}
Due to the uncertainties about the physical nature of the time-scale that normalizes the effects of rotation, the choice of such time-scale is somewhat arbitrary. For that reason we explore other alternatives for this normalizing time-scale, which have been measured within the cylinder defined by $R_{90}$ and $z_{90}$. These additional normalizing time-scales studied are shown in Figure \ref{fig:norm}. In the left panel we compare the mass-weighted averages of $\Omega$ and free-fall time $t_{\rm ff}^{\rm mid}$ at the mid-plane for different radii in the same disk, where $t_{\rm ff}$ is defined by $(G\rho_{s,g})^{-1/2}$ and the mid-plane is defined by the cell which has the maximum density along the z-axis. To compute this time-scale we consider the contribution of both gas and stars for the matter density $\rho_{s,g}$, resulting in the function:
\begin{equation}
\epsilon = 0.18_{-0.01}^{+0.01}e^{(-1.22\pm 0.03)\Omega t_{\rm ff}^{\rm mid}}
\end{equation}
 Although it seems to do a good job in normalizing the effects of rotation, there are still some theoretical issues with this physical parameter which question its nature. First, this time-scale represents the collapse of a spherical region of constant density which does not correspond to this case. Second, although stars affect the stability of a gaseous disk \citep{Rafikov_01}, they do not collapse with the gas, which is assumed to derive $t_{\rm ff}\propto (G\rho_{s,g})^{-1/2}$.

A second time-scale can be proposed by considering our numerical limitations. In order to form a star, a given gas cell must be collapsing due to self-gravity. We approach this problem by invoking the 3-dimensional Jeans instability, where any perturbation greater than the Jeans length $\lambda _J=c_s\sqrt{\frac{\pi}{G\rho}}$ will be gravitationally unstable and will trigger collapse. In the case of our simulations there is a critical length that corresponds to the spatial resolution $\Delta x$. Then by imposing $\Delta x = \lambda _J$ we find a second time-scale $(G\rho)^{-1/2} = \Delta x/\sqrt{\pi}c_s$. It should be mentioned that for a fixed $c_s$ this time-scale is an effective time delay imposed by our numerical recipes which depend on $\Delta x$. The effects of this time-scale are shown in the right panel of Figure \ref{fig:norm} and the best fit is given by:
\begin{equation}
\epsilon=0.23_{-0.01}^{+0.01}e^{(-2.43\pm 0.05)\Omega \Delta x/\sqrt{\pi}c_s}
\end{equation}
Although it is a natural time-scale of the simulations, its numerical nature and the non-sphericity of our systems put this time as a normalizing time-scale into doubt, requiring additional simulations of identical objects to depict its effects.

\section{SUMMARY AND CONCLUSIONS}
\label{sec:conclusions}

In this work we present simulations of spiral and starburst galaxies to study star formation. We find that their evolution is properly described by the relation proposed by \citet{Escala_15} which also suppresses the scatter due to inclination effects. Additionally, we find that the galactic rotation of disks reduces the efficiency of star formation and delays the equilibrium evolution phase. Finally we find a unique star formation law which incorporates the effects of galactic rotation. \\

The simulations were performed using the AMR Enzo code \citep{Bryan_14} employing the ZEUS hydro-solver \citep{Stone_92} to compute the hydrodynamics. We study four rotationally supported nuclear disks of starburst galaxies, with parameters $R \simeq 300$ pc, $M_{\rm gas}=4\times 10^8 M_{\odot}$ and a resolution of $\sim 2$pc, embedded in an isothermal sphere with a dispersion velocity of $\sigma=$100, 130, 190 and 220 km/s, and two rotationally supported spiral galaxies, with a radial length scale of $\rm R \simeq  3.5$ kpc, $M_{\rm gas}=10^{10} \msun$ and a resolution of $\sim 40$ pc, embedded in a potential generated by stars (Miyamoto-Nagai profile) and dark matter (NFW profile), where ($M_{\rm star}$,$M_{\rm DM}$)$\in\{ (10^{10}\msun,10^{10}\msun),(10^{11}\msun,10^{11}\msun)\}$.\\

The whole set of simulations shows to be well represented by two regimes as found empirically by \citet{Daddi_10} starburst galaxies lie along the starburst-like objects, and spiral galaxies lie in the normal star formation regime, each one of which follows a KS, $\ssfr\propto \sgas^N$, law with a $N =1.4$ index in concordance with observations of \citet{Kennicutt_98}). Although individual simulations show a good agreement with a SE law, $\ssfr\propto \sgas \Omega$, its functional form is not adequate since its efficiency seems to be anti-correlated with angular velocity. By introducing the effects of the inclination with respect to the line of sight we find that each point is shifted by a vector with a slope $\sim 1$ in the KS and SE plots (Fig. \ref{fig:disp3}), which translates into different and similar efficiencies in the KS and Silk plots respectively.\\

We test the dimensionally homogeneous equation $\ssfr = \epsilon \sqrt{G/L} \sgas ^{3/2}$, proposed by \citet{Escala_15}, where $L$ is the length scale along the line of sight. The temporal evolution of each individual simulation shows to be in agreement with this equation and we found that the efficiency under this prescription appears to be inclination-invariant, which means that a single object will always lie on the same star formation curve independently of the observer (Fig. \ref{fig:disp3}). \\

Studying each galaxy during the phase of star formation that is not affected by their initial conditions and numerical parameters, we find that the remaining scatter of the \citet{Escala_15} relation is parameterized by the angular velocity $\Omega$. In particular, we find that the star formation efficiency is diminished by an increasing $\Omega$. Furthermore, the star formation efficiency is well represented by an exponential decreasing function of $\Omega \tff ^{\rm ini}$ where $\Omega$ is the characteristic angular velocity and $\tff ^{\rm ini}$ is the initial free-fall time. This leads to a unique galactic star formation law, which incorporates the effects of galactic rotation, of the form

\begin{equation}
\label{eq:work}
\ssfr = 0.27e^{-1.91\Omega \tff ^{\rm ini}} \sqrt{\frac{G}{L}}\sgas ^{3/2}
\end{equation}
Notably, the exponential function of the efficiency has already been found in previous simulations \citep{Li_05,Padoan_12}. Finally, our recipe can be included in cosmological simulations where the star formation treatment relies on global properties due to their limited resolution.

\acknowledgments The simulations were performed using the HPC clusters Leftraru (ECM-02), the Geryon2 cluster (PFB06, QUIMAL 130008 and Fondequip AIC-57) and the Docorozco cluster from the Departamento de Astronom\'ia at Universidad de Chile. The analysis and plots were carried out with the publicly available tool yt \citep{Turk_11}. We are very grateful to Guillermo Blanc for the discussions and comments. We also thank the anonymous referee for fruitful suggestions and comments. JU acknowledges support from Programa Nacional de Becas de Postgrado, CONICYT (grant D-21140839). AE acknowledges partial support from the Center for Astrophysics and Associated Technologies CATA (PFB06), Anillo de Ciencia y Tecnolog\'ia (Project ACT1101) and Proyecto Regular Fondecyt (grant 1130458).

\bibliography{bibliography}

\begin{thebibliography}{}
\expandafter\ifx\csname natexlab\endcsname\relax\def\natexlab#1{#1}\fi

\bibitem[{{Becerra} \& {Escala}(2014)}]{Becerra_14}
{Becerra}, F., \& {Escala}, A. 2014, \apj, 786, 56

\bibitem[{{Berger} \& {Colella}(1989)}]{Berger_89}
{Berger}, M.~J., \& {Colella}, P. 1989, \jcoph, 82, 64

\bibitem[{{Berta} {et~al.}(2008){Berta}, {Jimenez}, {Heavens}, \&
  {Panter}}]{Berta_08}
{Berta}, Z.~K., {Jimenez}, R., {Heavens}, A.~F., \& {Panter}, B. 2008, \mnras,
  391, 197

\bibitem[{{Bigiel} {et~al.}(2008){Bigiel}, {Leroy}, {Walter}, {Brinks}, {de
  Blok}, {Madore}, \& {Thornley}}]{Bigiel_08}
{Bigiel}, F., {Leroy}, A., {Walter}, F., {et~al.} 2008, \aj, 136, 2846

\bibitem[{{Binney} \& {Tremaine}(2008)}]{Binney_Tremaine_08}
{Binney}, J., \& {Tremaine}, S. 2008, {Galactic Dynamics: Second Edition}
  (Princeton University Press)

\bibitem[{{Bryan} {et~al.}(2014){Bryan}, {Norman}, {O'Shea}, {Abel}, {Wise},
  {Turk}, {Reynolds}, {Collins}, {Wang}, {Skillman}, {Smith}, {Harkness},
  {Bordner}, {Kim}, {Kuhlen}, {Xu}, {Goldbaum}, {Hummels}, {Kritsuk}, {Tasker},
  {Skory}, {Simpson}, {Hahn}, {Oishi}, {So}, {Zhao}, {Cen}, {Li}, \& {Enzo
  Collaboration}}]{Bryan_14}
{Bryan}, G.~L., {Norman}, M.~L., {O'Shea}, B.~W., {et~al.} 2014, \apjs, 211, 19

\bibitem[{{Cen} \& {Ostriker}(1992)}]{Cen_Ostriker_92}
{Cen}, R., \& {Ostriker}, J.~P. 1992, \apjl, 399, L113

\bibitem[{{Ceverino} \& {Klypin}(2009)}]{Ceverino_09}
{Ceverino}, D., \& {Klypin}, A. 2009, \apj, 695, 292

\bibitem[{{Daddi} {et~al.}(2010){Daddi}, {Elbaz}, {Walter}, {Bournaud},
  {Salmi}, {Carilli}, {Dannerbauer}, {Dickinson}, {Monaco}, \&
  {Riechers}}]{Daddi_10}
{Daddi}, E., {Elbaz}, D., {Walter}, F., {et~al.} 2010, \apjl, 714, L118

\bibitem[{{Davis} {et~al.}(2014){Davis}, {Young}, {Crocker}, {Bureau}, {Blitz},
  {Alatalo}, {Emsellem}, {Naab}, {Bayet}, {Bois}, {Bournaud}, {Cappellari},
  {Davies}, {de Zeeuw}, {Duc}, {Khochfar}, {Krajnovi{\'c}}, {Kuntschner},
  {McDermid}, {Morganti}, {Oosterloo}, {Sarzi}, {Scott}, {Serra}, \&
  {Weijmans}}]{Davis_14}
{Davis}, T.~A., {Young}, L.~M., {Crocker}, A.~F., {et~al.} 2014, \mnras, 444,
  3427

\bibitem[{{Dobbs} {et~al.}(2015){Dobbs}, {Pringle}, \&
  {Duarte-Cabral}}]{Dobbs_15}
{Dobbs}, C.~L., {Pringle}, J.~E., \& {Duarte-Cabral}, A. 2015, \mnras, 446,
  3608

\bibitem[{{Elmegreen}(1997)}]{Elmegreen_97}
{Elmegreen}, B.~G. 1997, in \rmxac, Vol.~6, Revista Mexicana de Astronomia y
  Astrofisica Conference Series, ed. J.~{Franco}, R.~{Terlevich}, \&
  A.~{Serrano}, 165

\bibitem[{{Escala}(2015)}]{Escala_15}
{Escala}, A. 2015, \apj, 804, 54

\bibitem[{{Escala} \& {Larson}(2008)}]{Escala_08}
{Escala}, A., \& {Larson}, R.~B. 2008, \apjl, 685, L31

\bibitem[{{Genzel} {et~al.}(2010){Genzel}, {Tacconi}, {Gracia-Carpio},
  {Sternberg}, {Cooper}, {Shapiro}, {Bolatto}, {Bouch{\'e}}, {Bournaud},
  {Burkert}, {Combes}, {Comerford}, {Cox}, {Davis}, {Schreiber},
  {Garcia-Burillo}, {Lutz}, {Naab}, {Neri}, {Omont}, {Shapley}, \&
  {Weiner}}]{Genzel_10}
{Genzel}, R., {Tacconi}, L.~J., {Gracia-Carpio}, J., {et~al.} 2010, \mnras,
  407, 2091

\bibitem[{{Kennicutt}(1998)}]{Kennicutt_98}
{Kennicutt}, Jr., R.~C. 1998, \apj, 498, 541

\bibitem[{{Krumholz} {et~al.}(2012){Krumholz}, {Dekel}, \&
  {McKee}}]{Krumholz_12}
{Krumholz}, M.~R., {Dekel}, A., \& {McKee}, C.~F. 2012, \apj, 745, 69

\bibitem[{{Lada} {et~al.}(2010){Lada}, {Lombardi}, \& {Alves}}]{Lada_10}
{Lada}, C.~J., {Lombardi}, M., \& {Alves}, J.~F. 2010, \apj, 724, 687

\bibitem[{{Leroy} {et~al.}(2008){Leroy}, {Walter}, {Brinks}, {Bigiel}, {de
  Blok}, {Madore}, \& {Thornley}}]{Leroy_08}
{Leroy}, A.~K., {Walter}, F., {Brinks}, E., {et~al.} 2008, \aj, 136, 2782

\bibitem[{{Li} {et~al.}(2005){Li}, {Mac Low}, \& {Klessen}}]{Li_05}
{Li}, Y., {Mac Low}, M.-M., \& {Klessen}, R.~S. 2005, \apjl, 620, L19

\bibitem[{{Madore}(1977)}]{Madore_77}
{Madore}, B.~F. 1977, \mnras, 178, 1

\bibitem[{{Martin} \& {Kennicutt}(2001)}]{Martin_01}
{Martin}, C.~L., \& {Kennicutt}, Jr., R.~C. 2001, \apj, 555, 301

\bibitem[{{McKee} \& {Ostriker}(2007)}]{McKee_07}
{McKee}, C.~F., \& {Ostriker}, E.~C. 2007, \araa, 45, 565

\bibitem[{{Miyamoto} \& {Nagai}(1975)}]{Miyamoto_Nagai_75}
{Miyamoto}, M., \& {Nagai}, R. 1975, \pasj, 27, 533

\bibitem[{{Navarro} {et~al.}(1997){Navarro}, {Frenk}, \& {White}}]{NFW97}
{Navarro}, J.~F., {Frenk}, C.~S., \& {White}, S.~D.~M. 1997, \apj, 490, 493

\bibitem[{{Obreschkow} {et~al.}(2015){Obreschkow}, {Glazebrook}, {Bassett},
  {Fisher}, {Abraham}, {Wisnioski}, {Green}, {McGregor}, {Damjanov}, {Popping},
  \& {J{\o}rgensen}}]{Obreschkow_15}
{Obreschkow}, D., {Glazebrook}, K., {Bassett}, R., {et~al.} 2015, \apj, 815, 97

\bibitem[{{Padoan} {et~al.}(2014){Padoan}, {Federrath}, {Chabrier}, {Evans},
  {Johnstone}, {J{\o}rgensen}, {McKee}, \& {Nordlund}}]{Padoan_14}
{Padoan}, P., {Federrath}, C., {Chabrier}, G., {et~al.} 2014, \prsap, 77

\bibitem[{{Padoan} {et~al.}(2012){Padoan}, {Haugb{\o}lle}, \&
  {Nordlund}}]{Padoan_12}
{Padoan}, P., {Haugb{\o}lle}, T., \& {Nordlund}, {\AA}. 2012, \apjl, 759, L27

\bibitem[{{Rafikov}(2001)}]{Rafikov_01}
{Rafikov}, R.~R. 2001, \mnras, 323, 445

\bibitem[{{Rosen} \& {Bregman}(1995)}]{Rosen_95}
{Rosen}, A., \& {Bregman}, J.~N. 1995, \apj, 440, 634

\bibitem[{{Roychowdhury} {et~al.}(2015){Roychowdhury}, {Huang}, {Kauffmann},
  {Wang}, \& {Chengalur}}]{Roy_15}
{Roychowdhury}, S., {Huang}, M.-L., {Kauffmann}, G., {Wang}, J., \&
  {Chengalur}, J.~N. 2015, \mnras, 449, 3700

\bibitem[{{Sarazin} \& {White}(1987)}]{Sarazin_87}
{Sarazin}, C.~L., \& {White}, III, R.~E. 1987, \apj, 320, 32

\bibitem[{{Schmidt}(1959)}]{Schmidt_59}
{Schmidt}, M. 1959, \apj, 129, 243

\bibitem[{{Seigar}(2005)}]{Seigar_05}
{Seigar}, M.~S. 2005, \mnras, 361, L20

\bibitem[{{Semenov} {et~al.}(2016){Semenov}, {Kravtsov}, \&
  {Gnedin}}]{Semenov_15}
{Semenov}, V.~A., {Kravtsov}, A.~V., \& {Gnedin}, N.~Y. 2016, \apj, 826, 200

\bibitem[{{Silk}(1997)}]{Silk_97}
{Silk}, J. 1997, \apj, 481, 703

\bibitem[{{Stone} \& {Norman}(1992)}]{Stone_92}
{Stone}, J.~M., \& {Norman}, M.~L. 1992, \apjs, 80, 753

\bibitem[{{Suwannajak} {et~al.}(2014){Suwannajak}, {Tan}, \&
  {Leroy}}]{Suwannajak_14}
{Suwannajak}, C., {Tan}, J.~C., \& {Leroy}, A.~K. 2014, \apj, 787, 68

\bibitem[{{Tasker} \& {Tan}(2009)}]{Tasker_09}
{Tasker}, E.~J., \& {Tan}, J.~C. 2009, \apj, 700, 358

\bibitem[{{Toomre}(1964)}]{Toomre_64}
{Toomre}, A. 1964, \apj, 139, 1217

\bibitem[{{Truelove} {et~al.}(1997){Truelove}, {Klein}, {McKee}, {Holliman},
  {Howell}, \& {Greenough}}]{Truelove_97}
{Truelove}, J.~K., {Klein}, R.~I., {McKee}, C.~F., {et~al.} 1997, \apjl, 489,
  L179

\bibitem[{{Turk} {et~al.}(2011){Turk}, {Smith}, {Oishi}, {Skory}, {Skillman},
  {Abel}, \& {Norman}}]{Turk_11}
{Turk}, M.~J., {Smith}, B.~D., {Oishi}, J.~S., {et~al.} 2011, \apjs, 192, 9

\bibitem[{{Weidner} {et~al.}(2010){Weidner}, {Bonnell}, \&
  {Zinnecker}}]{Weidner_10}
{Weidner}, C., {Bonnell}, I.~A., \& {Zinnecker}, H. 2010, \apj, 724, 1503

\end{thebibliography}

\begin{appendix}
\section{RADIAL PROFILES}
\label{sec:profiles}
\begin{figure*}[ht!]
	\begin{center}
	\includegraphics[width=1.0\textwidth ,natwidth=400,natheight=600]{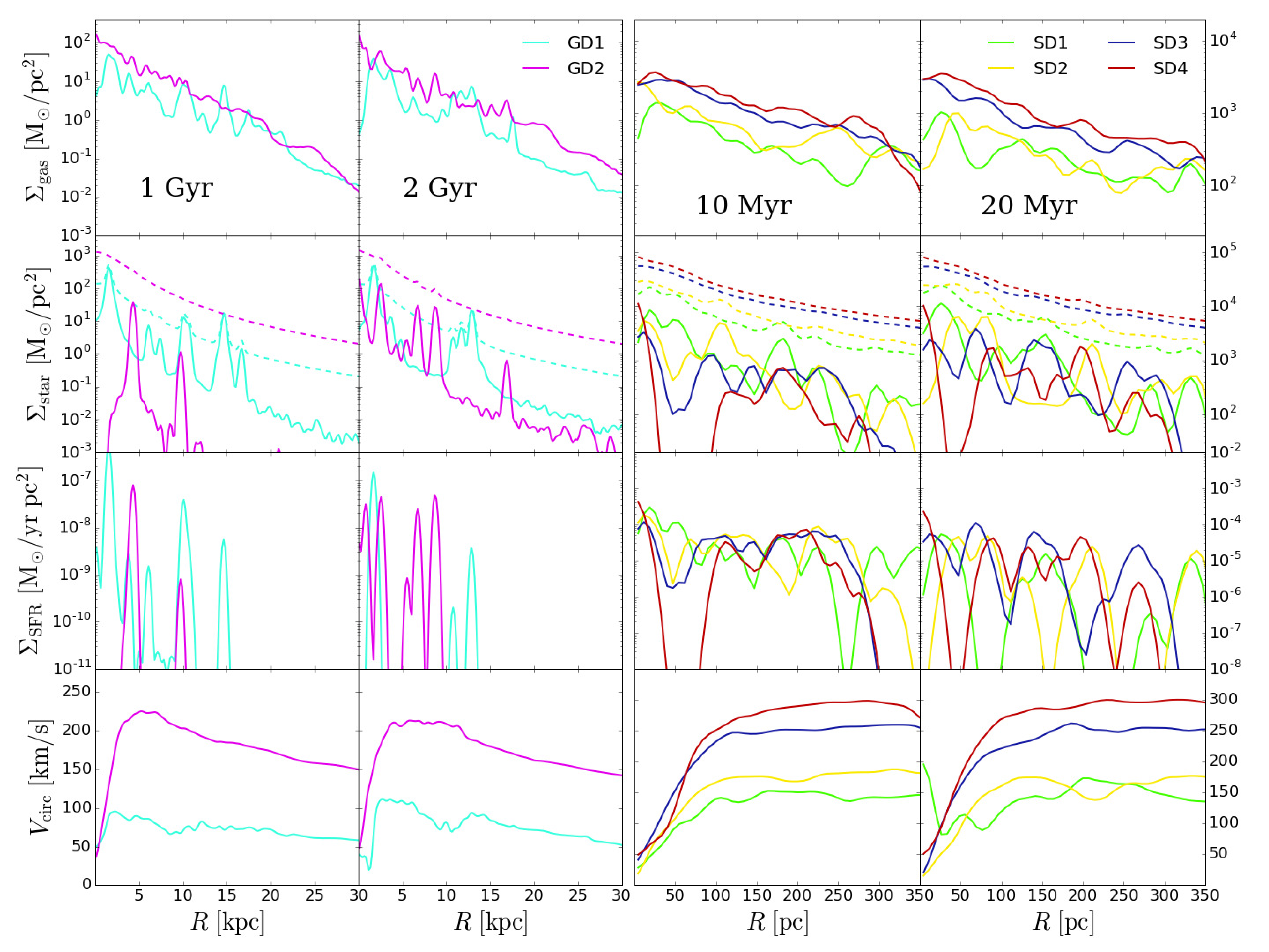}
	\caption{Radial profiles of $\sgas$, $\sstar$, $\ssfr$ and $V_{\rm circ}$: Left: radial profiles for spiral galaxies at 1 and 2 Gyr. Right: radial profiles for starburst galaxies. At the first column of each panel, runs GD1 and SD1 show clearly a fast gas depletion and high SFR despite their lower gas content. The second row shows the surface density of the created star particles (solid line) and the total stellar density (dashed line) that considers the stellar external potentials. The regions with current star formation notably changes the current $\sstar$ of the created star particles.}
	\label{fig:rprofiles}
	\end{center}
	\end{figure*}

Figure \ref{fig:rprofiles} shows radial profiles of surface gas density $\sgas$, surface stellar density $\sstar$, SFR per unit area $\ssfr$ and circular velocity $V_{\rm circ}$. These profiles are computed at times 1 and 2 Gyr for spiral galaxies and 10 and 20 Myr for starburst galaxies. The bin width $\Delta r$ is chosen to be four times the maximum spatial resolution for each simulation which corresponds to the smallest resolve Jeans length \citep{Truelove_97}. The quantities $\sgas$ and $\sstar$ are computed as the total mass of the respective component within a radial bin divided by the ring area $2\pi r\Delta r$. We do the same to compute $\ssfr$ but only consider the stars formed within 100 Myr for spiral galaxies and 1 Myr for starburst galaxies. For $V_{\rm circ}$ we take the median value of the circular velocity at each radius since the median is less sensitive to negative values of $V_{\rm circ}$ associated with the local rotational support of the collapsing structures.\\

\bigskip
\section{CHARACTERISTIC ANGULAR VELOCITY}
\label{sec:Omegas}
In this section we review different estimates of the characteristic angular velocity and test how each of these values changes our results.\\

The three most basic estimates for a characteristic value are the mean, mass-weighted average, and median values of the angular velocity which we denote as $\overline{\Omega}$, $\overline{\Omega}_{\rm M}$ and $\Omega_{\rm 50}$ respectively. It is important to point out that such estimates are meaningful for systems whose mass distribution and angular velocity profiles are similar.

Additionally, we can define other estimates which are based on the mass distribution or the process of star formation itself. As done in the main text we can define an angular velocity defined by the angular momentum  and moment of inertia in the $z$-direction, $\Omega_{\rm J}$. For a characteristic value based on star formation we consider two possibilities. First, we can measure $\Omega$ at the radius where $\sgas(R)$ is equal to the global $\sgas^{*}=M(R<R_{90})/\pi R_{90}^2$ used to derive the relations. We denote this value as $\Omega_{\Sigma}$ which is given by
\begin{equation}
\Omega_{\Sigma}=\Omega(R_{\Sigma}) \qquad \text{with} \qquad \sgas(R_{\Sigma})=\sgas^*
\end{equation}
Second, we can assume that at first order the SFR per unit area is given by the relation $\ssfr(R) = \epsilon \sgas^{1.5}(R)$. Then for an axisymmetric disk the global value of $\ssfr$ is
\begin{equation}
\overline{\Sigma}_{\rm SFR}=\frac{\int \ssfr dS}{\int dS}=\epsilon \frac{\int \sgas^{1.5} R dR}{\int R dR}
\end{equation}
There is a radius $R_{\star}$ at which 
\begin{equation}
\sgas(R_{\star}) = \left(\frac{\int \sgas^{1.5} R dR}{\int R dR}\right)^{2/3}
\end{equation}
corresponding to the radius at which $\overline{\Sigma}_{\rm SFR}=\ssfr$. We measure $\Omega$ at $R=R_{\star}$ and we denote this value as $\Omega_{\star}$.

To compute $\Omega_{\Sigma}$ and $\Omega_{\star}$ we use the radial profile of $\sgas$ as defined in section \ref{sec:profiles}.\\

Finally we can estimate the angular velocity at a fixed radius based on the radial profile of the circular velocity. We see in Figure \ref{fig:rprofiles} that the slope of $V_{\rm circ}$ changes at $R\simeq$ 5 kpc for spiral galaxies and at $R\simeq$ 150 pc for starburst galaxies. Measuring $\Omega$ at a fixed radius considerably decreases its uncertainty. We denote this quantity as $\Omega _{\rm V}$.\\

Table \ref{table:Omegatable} shows the fit parameters for the different values of a characteristic $\Omega$ mentioned here. The parameters are fitted assuming a star formation relation of the form:
\begin{equation}
\ssfr = \alpha \exp(-\beta \Omega t_{\rm ff}) \sqrt{\frac{G}{L}} \sgas^{1.5}
\end{equation}

\begin{deluxetable}{cccc}
\tablecaption{Alternatives for $\Omega$ \label{table:Omegatable}}
\tabletypesize{\footnotesize}
\tablewidth{0pt}
\tablehead{
\colhead{$\Omega$} & \colhead{$\alpha$}&\colhead{$\beta$} & \colhead{Scatter}\\
\colhead{Name} & \colhead{(dex)}&\colhead{Name} & \colhead{(dex)}}
\startdata 
$\overline{\Omega}$ 			& 0.289 $\pm$ 0.046 &  1.734 $\pm$ 0.177 & 0.209 \\ 
$\overline{\Omega}_{\rm M}$ 	& 0.294 $\pm$ 0.078 &  1.359 $\pm$ 0.217 & 0.235 \\ 
$\overline{\Omega}_{\rm 50}$	& 0.251 $\pm$ 0.041 &  1.908 $\pm$ 0.208 & 0.212 \\ 
$\Omega_{\Sigma}$ 			& 0.356 $\pm$ 0.060 &  1.777 $\pm$ 0.162 & 0.265 \\ 
$\Omega_{\star}$ 			& 0.294 $\pm$ 0.051 &  1.497 $\pm$ 0.155 & 0.263 \\ 
$\Omega_{\rm V}$ 			& 0.461 $\pm$ 0.089 &  1.950 $\pm$ 0.181 & 0.202 \\ 
\hline
$\Omega_{\rm J}$              & 0.272 $\pm$  0.051 &  1.908 $\pm$ 0.222 & 0.206 \\
\enddata
\tablecomments{Fit parameters to the function $\epsilon = \alpha \exp(-\beta \Omega t_{\rm ff})$ for different estimations of the angular velocity $\Omega$. The efficiency $\epsilon$ corresponds to the efficiency of the relation proposed by \citet{Escala_15}. Our result is shown in the last row, where $\Omega_{\rm J}$ is the angular velocity based on angular momentum.}
\end{deluxetable}

Our choice for the angular velocity $\Omega_{\rm J}$ shows an amplitude $\alpha$ consistent with most of the results for different values of $\Omega$. Additionally, $\Omega_{\rm J}$ shows a slightly smaller scatter than its counterparts. The value $\Omega_{\rm V}$ results in a smaller scatter but also shows the largest deviation in $\alpha$ with respect to the other values.

\end{appendix}

\end{document}